\documentclass[12pt,final] {article}

\usepackage{amsmath}
\usepackage{amsfonts}
\usepackage{amssymb}
\usepackage{amsthm}
\usepackage{mathtools}
\usepackage{dsfont}
\usepackage{mathrsfs}
\usepackage{graphicx}
\usepackage{color}
\usepackage[utf8]{inputenc}
\usepackage[margin=1in]{geometry}
\usepackage[titletoc]{appendix}
\usepackage{bm}
\usepackage{algorithm}
\usepackage{algpseudocode}
\usepackage{afterpage}
\usepackage{enumitem}
\usepackage{multirow}
\usepackage{booktabs}
\usepackage{subfigure}
\usepackage[authoryear,sort]{natbib}
\usepackage{hhline}
\usepackage{adjustbox}
\usepackage[tableposition=top]{caption} 
\usepackage{epic}
\usepackage{xr}
\numberwithin{equation}{subsection}

\newtheorem{assumption}{Assumption}
\newlength{\vslength}
\usepackage{hyperref}
\hypersetup{
    colorlinks=true,
    citecolor=blue,
    linkcolor=blue,
    filecolor=blue,      
    urlcolor=blue,
    pdftitle={Overleaf Example},
    pdfpagemode=FullScreen, 
    }

\newtheorem{theorem}{Theorem}[section]

\newtheorem{lemma}{Lemma}[section]

\newtheorem{proposition}{Proposition}[section]

\newtheorem{corollary}{Corollary}[section]

\theoremstyle{remark}

\newtheorem{example}{\bf Example}[section]

\newtheorem{definition}{\bf Definition}[section]
\renewenvironment{proof}{\noindent{\it Proof.}}{\qed}

\newcommand{\bc}{\begin{center}}
\newcommand{\ec}{\end{center}}
\newcommand{\be}{\begin{equation}}
\newcommand{\ee}{\end{equation}}
\newcommand{\ba}{\begin{array}}
\newcommand{\ea}{\end{array}}
\newcommand{\bean}{\setlength\arraycolsep{1pt}\begin{eqnarray*}}
\newcommand{\eean}{\end{eqnarray*}}
\newcommand{\bea}{\setlength\arraycolsep{1pt}\begin{eqnarray}}
\newcommand{\eea}{\end{eqnarray}}
\newcommand{\ben}{\begin{enumerate}}
\newcommand{\een}{\end{enumerate}}
\newcommand{\bed}{\begin{itemize}}
\newcommand{\eed}{\end{itemize}}

\def\log{\hbox{log}}

\def\boxit#1{\vbox{\hrule\hbox{\vrule\kern6pt
 \vbox{\kern6pt#1\kern6pt}\kern6pt\vrule}\hrule}}

\def\bse{\begin{eqnarray*}}
\def\ese{\end{eqnarray*}}
\def\be{\begin{eqnarray}}
\def\ee{\end{eqnarray}}
\def\bq{\begin{equation}}
\def\eq{\end{equation}}
\def\bse{\begin{eqnarray*}}
\def\ese{\end{eqnarray*}}

\def\boxit#1{\vbox{\hrule\hbox{\vrule\kern6pt
          \vbox{\kern6pt#1\kern6pt}\kern6pt\vrule}\hrule}}

\def\diag{\hbox{diag}}

\def\diag{\hbox{diag}}
\def\log{\hbox{log}}

\def\bse{\begin{eqnarray*}}
\def\ese{\end{eqnarray*}}
\def\be{\begin{eqnarray}}
\def\ee{\end{eqnarray}}
\def\bq{\begin{equation}}
\def\eq{\end{equation}}
\def\bse{\begin{eqnarray*}}
\def\ese{\end{eqnarray*}}

\newcommand{\uiota}             {\mbox{\boldmath$\uiota$}}

\setcitestyle{authoryear,open={(},close={)}} 
\usepackage{relsize}
\usepackage{geometry}
\usepackage{booktabs, multirow}

\begin{document}

\thispagestyle{empty}
\title{
    \vspace*{-9mm}
    Directed Cyclic Graph for Causal Discovery from Multivariate Functional Data}
\author{Saptarshi Roy,\ Raymond K.~W.~Wong,\ Yang Ni \\
    {\small\it {}Department of Statistics, Texas A$\&$M University}
}
\date{September 22, 2023}

\maketitle
\begin{abstract}
Discovering causal relationship using multivariate functional data has received a significant amount of attention very recently. In this article, we introduce a functional linear structural equation model for causal structure learning when the underlying graph involving the multivariate functions may have cycles. To enhance interpretability, our model involves a low-dimensional causal embedded space such that all the relevant causal information in the multivariate functional data is preserved in this lower-dimensional subspace. We prove that the proposed model is causally identifiable under standard assumptions that are often made in the causal discovery literature. To carry out inference of our model, we develop a fully Bayesian framework with suitable prior specifications and uncertainty quantification through posterior summaries. We illustrate the superior performance of our method over existing methods in terms of causal graph estimation through extensive simulation studies. We also demonstrate the proposed method using a brain EEG dataset. 
\end{abstract}

\section{Introduction}
   \paragraph{Motivation.}  
Multivariate functional data arise in many fields such as
biomedical research \citep{wei_Li, chiou2016pairwise}, environmental science \citep{kortestapff2022multivariate}, finance \citep{kowal2017bayesian}, plant science \citep{wong2019partially, park2022crop}, and sport science \citep{volkmann2021multivariate} where multiple variables are measured over time or other domains. The increasing availability of functional data in these fields provides us with great opportunities to discover causal relationships among random functions for the better understanding of complex systems, which is helpful for various machine learning and statistics tasks such as representation learning \citep{causalRepresentLearn}, fairness \citep{Tang_2023}, transfer learning \citep{transferlearning}, and reinforcement learning \citep{zeng2023survey}. One motivating example is electroencephalography (EEG) where electrical activity from the brain is recorded non-invasively from electrode channels by placing them on the scalp or directly on the surface of the brain. Given its continuous nature and the short time separation between the adjacent measuring points, it is natural to treat the data at each brain location/region as a function over time. A relevant scientific goal is to estimate brain effective connectivity among different regions, which will potentially allow us to make better decisions, design more effective interventions, and avoid unintended consequences. However, existing structural equation model (SEM) based causal discovery methods assume acyclic relationships among the random functions by imposing a directed acyclic graph (DAG) structure, which may be too restrictive for many real applications. For example, there are strong indications that in brain effective connectivity studies, due to reciprocal polysynaptic connections, the brain regions are far from exhibiting acyclicity \citep{cycles1, cycles2}, and that in genetic pathways, due to the presence of multiple upstream regulators and downstream targets for every signaling component, feedback loops/directed cycles are regular motifs \citep{brandman2008feedback}. Thus, in light of the prevalence of cycles in complex systems, it is desirable to have a flexible model for causal discovery among random functions that can account for such cyclic causal structures.

\paragraph{Challenges.} Causal discovery for multivariate functional data in the presence of cycles is an inherently difficult problem that is not yet well understood. We highlight three prominent challenges. (i) Functional data are infinite-dimensional in nature. It may so happen that the low-frequency spectrum of one curve might causally influence the high-frequency spectrum of another curve. This demands identification of pertinent features that can be used to create a finite-dimensional representation of the data, which is easier to work with and analyze. However, the challenge is that we may not know \emph{a priori} what the relevant features are when dealing with infinite-dimensional objects.
Blind adoption of standard (non-causal-adaptive) low-dimensional features can lead to errors or inaccuracies. (ii) Although the identifiability of causal models for multivariate functional data in the absence of cycles has been established in recent works \citep{fBN, fSEM}, showing identifiability of causal models from multivariate data, let alone multivariate functions, is still a challenging and complex task in cases where causal relationships are obscured by the presence of cycles. (iii) It is common that functional data are only observed over discrete time points with additional noises.
Such incomplete and noisy observations of the functions add another layer of difficulty in probing the causal relationships of interest.

\paragraph{Related work.} Causal discovery from multivariate functional data has been studied by a few recent works \citep{fBN, fSEM, pmlr-v186-yang22a}, which have already shown some promising results in discovering causality in, e.g., EEG data and longitudinal medical record data. However, all of them are limited to DAGs, which do not allow inference of cyclic causality. While there has been a surge of research on causal discovery methods for scalar random variables in the presence of feedback loops/cycles over the last few decades \citep{ccd, Lacerda, CANM_Mooij, HyttinenEH12, CDHD, CCD_equillibrium, ConstrainBasedCyclic_Mooij, CHOD}, none of these approaches have been extended to discovering causal dependencies among random functions in multivariate settings. Therefore, how to handle cyclic causal relationships among multivariate functional data while addressing the aforementioned challenges remains a largely unsolved problem.

\paragraph{Contributions.}In this paper, we propose  
an operator-based non-recursive linear structural equation based novel causal discovery framework that identifies causal relationships among functional objects in the presence of cycles and additional measurement/sampling noises. Our major contribution is four-fold. 
\begin{enumerate}
    \item  We consider a causal embedding of the functional nodes into a lower-dimensional space for dimension reduction that adapts to causal relationships. 
    \item We prove that the causal graph of the proposed model is uniquely identifiable under standard causal assumptions. 
    \item  We capture within-function dependencies using a data-driven selection of orthonormal basis that is both interpretable and computationally efficient. 
    \item  To perform inference and uncertainty quantification from finite-sample data, we adopt a fully Bayesian hierarchical formulation with carefully selected prior distributions. Posterior inference is performed using Markov chain Monte Carlo (MCMC). We demonstrate the effectiveness of the proposed method in identifying causal structure and key parameters through simulation studies and apply the framework to the analysis of brain EEG data, illustrating its real-world applicability. Codes will be made available on the project's website on \href{https://github.com/roys8001}{Github}. 
\end{enumerate}

\section{Model Definition and causal identifiability}
   \subsection{Notations} \label{sec:2.1}
Let $[p] = \{1, \dots, p\}$ for any positive integer $p$. A causal directed cyclic graph (DCG) is a graph $\mathcal{G} = (\mathcal{V}, \mathcal{E})$, which consists of a set of vertices or nodes $\mathcal{V}=[p]$ representing a set of random variables and a set of directed edges $\mathcal{E}=\{\ell\to j|j,\ell\in \mathcal{V}\}$ representing the direct causal relationships among the random variables. In a DCG, we do not assume the graph to be acyclic. A causal DCG model is an ordered pair $(\mathcal{G}, \mathbb{P})$ where $\mathbb{P}$ is a joint probability distribution over $\mathcal{V}$ (more rigorously, the random variables that $\mathcal{V}$ represents) that satisfies conditional independence relationships encoded by the causal DCG $\mathcal{G}$. A simple directed cycle is a sequence of distinct vertices $\{v_1, \dots, v_k\}$ such that the induced subgraph by these vertices is $v_{1} \rightarrow \dots \rightarrow v_{k} \rightarrow v_{1}$.
For a vertex $j \in \mathcal{V}$, we use $\mathrm{pa}(j)$ to denote the set of parents (direct causes).

\subsection{Model framework}

Consider a multivariate stochastic process $\bm{Y}=(Y_1,\dots, Y_p)^\top$ where each $Y_{j}$ is defined on a compact domain $\mathcal{T}_{j}\subset \mathbb{R}$. Without loss of generality, we assume $\mathcal{T}_{1} = \dots = \mathcal{T}_{p}=[0,1]$. Suppose $Y_j\in\mathcal{H}_j$
where $\mathcal{H}_j$ is a Hilbert space of functions defined on $\mathcal{T}_{j}$. We let $\langle\cdot, \cdot\rangle$ denote the inner product of $\mathcal{H}_j$. 
We propose a causal model that captures the relationships among $Y_1,\dots, Y_p$.
 
Our proposed model considers an operator-based non-recursive linear structural equation model 
on the random functions $\bm{Y}$ as
\begin{equation}
\label{eqn:2.2.1}
    Y_j(\cdot) = \sum_{\ell\in\mathrm{pa}(j)} (\mathcal{B}_{j\ell} Y_\ell)(\cdot) + f_j(\cdot), \quad \forall j\in[p],
\end{equation}
where $\mathcal{B}_{j\ell}$ is a linear operator that maps $\mathcal{H}_\ell$ to $\mathcal{H}_j$, and $f_j\in\mathcal{H}_j$ is an exogenous stochastic process. Clearly, for any $j, \ell\in \mathcal{V}$ such that the edge $\ell \rightarrow j\in \mathcal{E}$, $\mathcal{B}_{j\ell}$ is not a null operator. Now by stacking the $p$ equations in (\ref{eqn:2.2.1}), we obtain 
\begin{equation}
\label{eqn:2.2.2}
   \bm{Y}= \mathfrak{B}\bm{Y} + \bm{f}, 
\end{equation}
 where $\mathfrak{B} = (\mathcal{B}_{j\ell})_{j,\ell = 1}^{p}$ is a matrix of operators and $\bm{f}=(f_1,\dots,f_p)^\top$ is a $p$-variate stochastic process. In DAGs, the causal effect matrix can be arranged into a lower block triangular structure given  a topological/causal ordering. But since our model allows for  cycles, we have no such restriction on the structure of the operator matrix $\mathfrak{B}$ except that $\mathcal{B}_{jj}, \forall j\in [p]$, is null, i.e., no self-loops.
 
Model \eqref{eqn:2.2.1} is infinite-dimensional and hence challenging to estimate and interpret.
To alleviate such difficulties, we consider a low-dimensional causal embedding structure.
Specifically, we assume that the causal relationships are preserved in an unknown low-dimensional subspace $\mathcal{D}_j$ of $\mathcal{H}_j$. Denote the dimension of $\mathcal{D}_j$ by $K_j$.
Let $\mathcal{P}_j$ and $\mathcal{Q}_j$
be the projection onto $\mathcal{D}_j$ and its orthogonal complement in $\mathcal{H}_j$ respectively.
We assume ${\mathcal{B}}_{j\ell}=\mathcal{P}_j\mathcal{B}_{j\ell}\mathcal{P}_\ell$, which implies that causal effects can be fully described within the low-dimensional subspaces $\{\mathcal{D}_j\}_{j=1}^p$.
As such, (\ref{eqn:2.2.1}) can be split into
\begin{align}
\label{eqn:2.2.3}
\mathcal{P}_j Y_j &= \sum_{\ell\in\mathrm{pa}(j)} {\mathcal{B}}_{j\ell} (\mathcal{P}_\ell Y_\ell) + \mathcal{P}_j f_j,\\
\mathcal{Q}_j Y_j &= \mathcal{Q}_jf_j.\nonumber
\end{align}

 \noindent We assume that $\mathcal{P}_{j}f_{j}$ and $\mathcal{Q}_{j}f_{j}$ are independent of each other. Now, by defining $\alpha_{j} = \mathcal{P}_{j}Y_{j}$ and $\epsilon_{j} = \mathcal{P}_{j}f_{j}, \forall j\in[p]$,
\eqref{eqn:2.2.3} can be compactly written as
\begin{equation}
\label{eqn:2.2.4}
    \bm{\alpha} = {\mathcal{B}}\bm{\alpha} + \bm{\epsilon},
\end{equation}
where $\bm{\alpha} = (\alpha_{1}, \dots, \alpha_{p})^{\top}$  and $\bm{\epsilon} = (\epsilon_{1}, \dots, \epsilon_{p})^{\top}$ with $\alpha_{j}, \epsilon_{j} \in \mathcal{D}_{j}, \forall j\in[p]$.

In practice, the random functions in $\bm{Y}$ can only be observed over a finite number of (input) locations, possibly with measurement errors.
 More specifically, for each random function $Y_j$, we observe $\{(t_{ju}, X_{ju})\}_{u=1}^{m_j}$, where $X_{ju}\in\mathbb{R}$ is the measurement of $Y_j$ at location $t_{ju}\in\mathcal{T}_j$ and $m_j$ is the number of measurements obtained from $Y_j$. Defining $\beta_{j} = \mathcal{Q}_{j}Y_{j}$, we consider the following measurement model:
\begin{align}
    X_{ju} &= Y_j(t_{ju})+ e_{ju}\nonumber\\ 
    \label{eqn:2.2.5}
   &= \alpha_{j}(t_{ju}) + \beta_{j}(t_{ju}) +  e_{ju},  \quad \forall u\in[m_{j}], j\in[p],
\end{align}
with independent noises $e_{ju} \sim N(0, \sigma_{j}), \forall u\in [m_{j}]$. 

More compactly, (\ref{eqn:2.2.5}) can be written as
\begin{align}
\label{eqn:2.2.6}
    \bm{X} = \bm{\alpha}(\bm{t}) + \bm{\beta}(\bm{t}) + \bm{e},
\end{align}
where   $\bm{X} = (\bm{X}^{\top}_{1}, \dots, \bm{X}^{\top}_{p})^{\top}$, $\bm{\alpha}(\bm{t}) = (\bm{\alpha}_{1}(\bm{t}_{1})^{\top}, \dots, \bm{\alpha}_{p}(\bm{t}_{p})^{\top})^{\top}$, $\bm{\beta}(\bm{t}) = (\bm{\beta}_{1}(\bm{t}_{1})^{\top}, \dots, \bm{\beta}_{p}(\bm{t}_{p})^{\top})^{\top}$ and $\bm{e} = (\bm{e}^{\top}_{1}, \dots, \bm{e}^{\top}_{p})^{\top}$ with $\bm{X}_{j} = (X_{j1}, \dots, X_{jm_{j}})^{\top}, \bm{\alpha}_{j}(\bm{t}_{j}) = (\alpha_{j}(t_{j1}), \dots, \alpha
_{j}(t_{jm_{j}}))^{\top}, \bm{\beta}_{j}(\bm{t}_{j}) = (\beta_{j}(t_{j1}), \dots, \beta
_{j}(t_{jm_{j}}))^{\top}$ and $\bm{e}_{j} = (e_{j1}, \dots, e_{jm_{j}})^{\top}$.

We call our proposed model, \textbf{FENCE}, which stands for '\textbf{F}unctional \textbf{E}mbedded \textbf{N}odes for \textbf{C}yclic causal \textbf{E}xploration', reflecting its purpose.

   \subsection{Causal identifiability} \label{sec:2.3}
In this section, we shall show that the graph structure of the proposed FENCE model is identifiable for functional data measured discretely with random noises under several causal assumptions. We start by defining causal identifiability and state our assumptions.
\begin{definition}(\textit{Causal Identifiability})
Suppose $\bm{Y}$ is a $p$-variate random function and $\bm{X}$ is the observed noisy version of $\bm{Y}$ given by (\ref{eqn:2.2.6}). Assume $\bm{X}$ follows FENCE model $\mathcal{S} = (\mathcal{G}, \mathbb{P})$ where $\mathcal{G}$ is the underlying graph and $\mathbb{P}$ is the joint distribution of $\bm{X}$ over $\mathcal{G}$. We say that $\mathcal{S}$ is causally identifiable from $\bm{X}$ if there does not exist any other $\mathcal{S}^{*} = (\mathcal{G}^{*}, \mathbb{P}^{*})$ with $\mathcal{G}^{*} \neq \mathcal{G}$ such that the joint distribution $\mathbb{P^{*}}$ on $\bm{X}$ induced by $\mathcal{G}^{*}$ is equivalent to $\mathbb{P}$ induced by $\mathcal{G}$.
\end{definition}

In other words, for a causal graph to be identifiable, there must not exist any other graph such that the joint distributions induced by the two different graphs are equivalent. Next, we list and discuss a few assumptions to establish the causal identifiability of the proposed model.

\begin{assumption}(Causal Sufficiency)\label{ass1}
The model $\mathcal{S} = (\mathcal{G}, \mathbb{P})$ is causally sufficient, i.e., there are no unmeasured confounders.
\end{assumption}
Assuming no unmeasured confounders keeps the causal discovery task more manageable especially for cyclic graphs with purely observational data.

\begin{assumption}(Disjoint Cycles)\label{ass2}
The cycles in $\mathcal{G}$ are disjoint, i.e., no two cycles in the graph have two nodes that are common to both. 
\end{assumption}
Assuming disjoint cycles induces a natural topological ordering and forms a directed acyclic hypergraph-like structure within the DCG. The same assumption was made in \cite{Lacerda}. 

\begin{assumption}(Stability)\label{ass3}
For the model $\mathcal{S}$, the moduli of the eigenvalues of the finite rank operator $\mathcal{B}$ are less than or equal to $1$, and none of the real eigenvalues are equal to $1$. 
\end{assumption}

According to \citealp{fisher}, the SEM in (\ref{eqn:2.2.4}) can be viewed as being in a state of equilibrium, where the finite rank operator $\mathcal{B}$ represents coefficients in a set of dynamical equations that describe a deterministic dynamical system observed over small time intervals as the time lag approaches zero. The eigenvalue conditions are deemed necessary and sufficient for the limiting behavior to hold, as argued by \citealp{fisher}. Such an assumption is widely adopted in e.g., econometrics, and \citealp{Lacerda} made this assumption as well.

\begin{assumption}(Non-Gaussianity)\label{ass4}
The exogenous variables have independent mixture of Gaussian distributions.
i.e., $\epsilon_{jk} \overset{\text{ind}}{\sim} \sum_{m = 1}^{M_{jk}} \pi_{jkm} \text{N}(\mu_{jkm}, \tau_{jkm})$ with $M_{jk} \geq 2$.
\end{assumption}

The assumption of non-Gaussianity on the exogenous variables has been proven useful in causal discovery as it induces model identifiability in the linear SEM framework \citep{HP, shimizu06a, Lacerda, spirtes2016}. Mixture of Gaussian can approximate any continuous distribution arbitrarily well given a sufficiently large number of mixture components \citep{titterington_85, MCLA2000, rossi}. It is also easy to sample, which facilitates our posterior inference.

\begin{assumption}\label{ass5}
(Non-causal dependency) We assume $\bm{\beta}(\bm{t}) = \bm{C}(\bm{t})\bm{\gamma}$, where $\bm{\gamma}$ represent another exogenous component of the model and $\bm{C}(\bm{t}) = diag(\bm{C}_{11}(\bm{t}_{1}), \dots, \bm{C}_{pp}(\bm{t}_{p}))$. Here $\bm{C}_{jj}(\bm{t}_j)$ is a mixing matrix that mixes the independent entires in $\bm{\gamma}$ to generate temporal dependence within the $j$-th block. We assume $\gamma_{jk} \overset{\text{ind}}{\sim} \sum_{m = 1}^{M_{jk}} \pi_{jkm}' \text{N}(\mu_{jkm}', \tau_{jkm}')$ with $M_{jk}\geq 1$. 
\end{assumption}

Since the model assumes that all causal information in $\bm{Y}$ is preserved in the lower-dimensional space $\mathcal{D}_{j}$ and not in its orthogonal complement, it is apparent that while each $\bm{\beta}_{j}(\bm{t}_{j})$ within a block can have temporal dependence, it is independent of $\bm{\beta}_{\ell}(\bm{t}_{\ell})$ when $j\neq \ell$ and $j, \ell \in [p]$.

For some basis $\{\phi_{jk}\}_{k=1}^{K_{j}}$ that spans the low-dimensional causal embedded space $\mathcal{D}_{j}$, $\alpha_{j}$ in (\ref{eqn:2.2.5}) can be further expanded by, $\alpha_{j}(t_{ju}) = \sum_{k = 1}^{K_{j}}\tilde{\alpha}_{jk}\phi_{jk}(t_{ju})$. Therefore (\ref{eqn:2.2.6}) can be written more compactly as 
\begin{equation}
    \bm{X} = \bm{\Phi}(\bm{t})\tilde{\bm{\alpha}} + \bm{\beta}(\bm{t}) + \bm{e},
    \label{eqn:2.2.7}
\end{equation}
where $\bm{\Phi}(\bm{t}) = \diag(\bm{\Phi}_{1}(\bm{t}_{1}), \dots, \bm{\Phi}_{p}(\bm{t}_{p}))$ with $\bm{\Phi}_{j}(\bm{t}_{j}) = (\phi_{jv}(t_{ju}))_{u=1, v=1}^{m_{j}, K_{j}}$.

\begin{assumption}\label{ass6}
(Sufficient sampling locations) The basis matrix $\bm{\Phi}(\bm{t})$ of size $\sum_{j=1}^{p}m_{j} \times \sum_{j=1}^{p}K_{j}$ has a full column rank. 
\end{assumption}

This assumption implies enough sampling locations, over which each random function $Y_j$ is observed, to capture all the causal information that $Y_j$ contains. 

Given these six assumptions, our main theorem establishes the causal identifiability of the proposed model.

\begin{theorem} \label{Th:1}
Under Assumptions \ref{ass1} - \ref{ass6}, $\mathcal{S} = (\mathcal{G}, \mathbb{P})$ is causally identifiable.
\end{theorem}

\begin{figure}[htb]
    \centering
    \includegraphics[width=10cm]{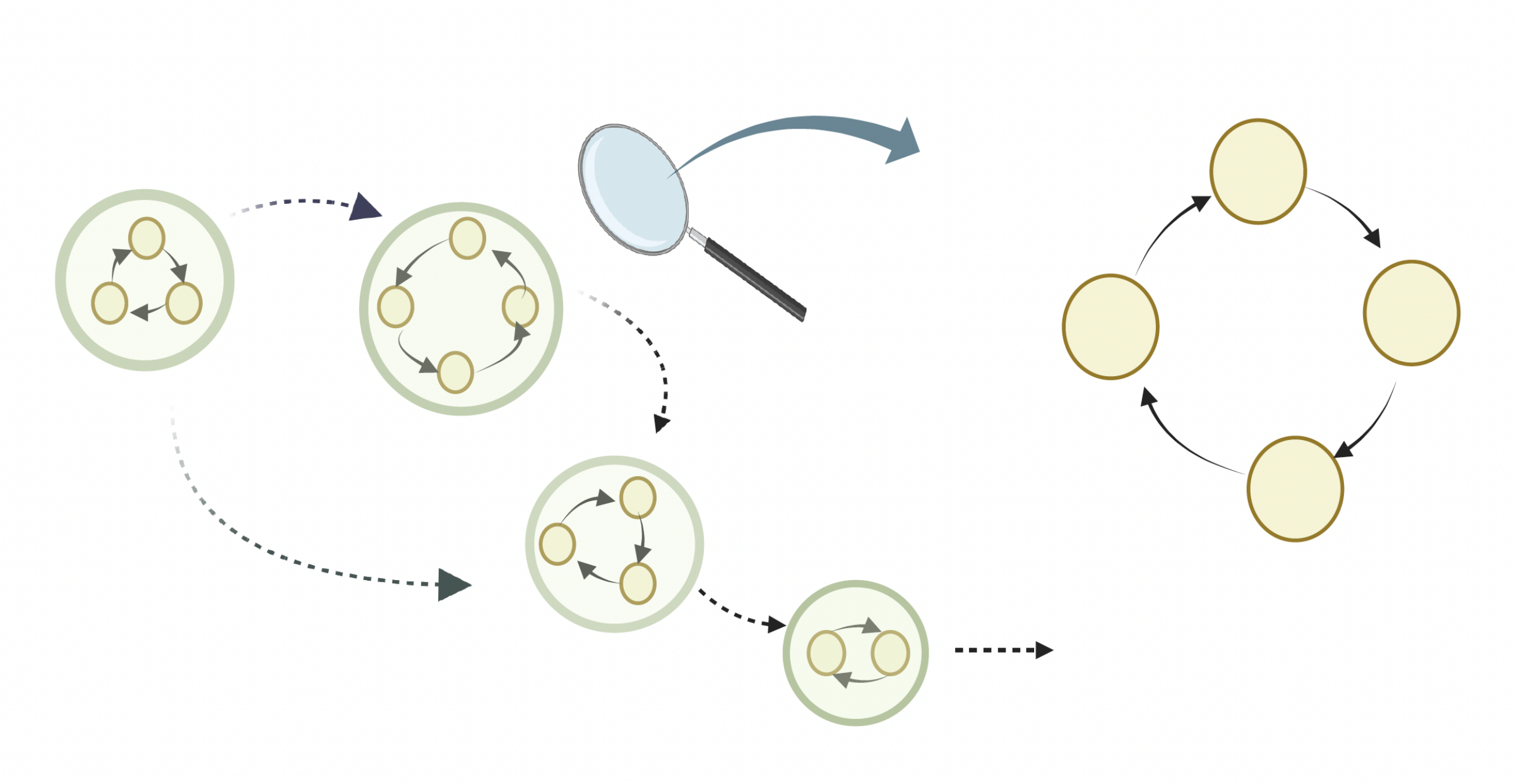}
    \caption{Two important components of causal identifiability proof: (I) identifiability of directed acyclic hypergraph induced by disjoint cycles, and (II) identifiability of each disjoint cycle.}
    \label{fig:1}
\end{figure}

The proof essentially involves two steps as shown in Figure \ref{fig:1}. On the left-hand side (LHS) of the diagram, we depict the hypergraph-like structure that emerges when assuming the existence of disjoint cycles (Assumption \ref{ass2}), whereas, on the right-hand side (RHS), we offer a magnified view of the hypernodes (nodes containing simple directed cycle). Our approach to proving causal identifiability progresses from the LHS to the RHS. That is, we first prove the identifiability of the hypergraph-like structure depicted on the LHS of Figure \ref{fig:1}, and then we proceed to establish the identifiability of each simple directed cycle within every hypernode in the hypergraph. The detailed exposition of the proof can be found in Section \ref{section:proof} of the Supplementary Materials. 

\section{Bayesian Model Formulation}
   
In this section, we will describe the inference procedure of the proposed model. A straightforward approach would be a two-step procedure where the first step performs functional principal component analysis on each function marginally to reduce the dimension, and then the second step learns causal structure based on the principal components. However, this simple approach has several disadvantages. First, the estimated functional principal components that explain the most variation of each individual function marginally may not optimally capture the cause-effect dependence relationships among different functions. Second, this procedure is unreliable since estimation uncertainty fails to propagate correctly from the first step to the second step. As such, we propose a fully Bayesian approach, which reduces the dimension of functional data adaptively for causal structure learning. 

\subsection{Model parameters}
Let $\bm{E} = (E_{j\ell})_{j,\ell = 1}^{p}$ denote the adjacency matrix where $E_{j\ell} = 1$ indicates the existence of a directed edge from node $\ell$ to node $j$, and 0 otherwise. Let $\{\phi_{k}\}_{k = 1}^{S}$ be a set of $S$ common unknown basis functions that approximate each random function $Y_{j}$ i.e., $Y_{j} = \sum_{k = 1}^{S}\tilde{\alpha}_{jk}\phi_{k}$ where $\{\tilde{\alpha}_{jk}\}_{k=1}^{S}$ denote the set of basis coefficients.  
Note that $\{\phi_{k}\}$ is not the basis for the lower-dimensional causal embedded subspace $\mathcal{D}_{j}$. However, we assume that the first $K_{j}$ of them actually spans $\mathcal{D}_{j}$ and our goal is to hunt for them through a properly designed inference procedure. Moreover, according to our assumption, we build our SEM on the first $K_{j}$ of the basis coefficients $ \tilde{\bm{\alpha}}_{j} = (\tilde{\alpha}_{j1}, \cdots, \tilde{\alpha}_{jK_{j}})^{\top}$. Defining $\bar{\bm{\alpha}}_{j} = (\tilde{\alpha}_{j,K_{j} + 1}, \cdots, \tilde{\alpha}_{jS})^{\top}$ with $\bar{\bm{\alpha}}_{j} = \bm{\gamma}_{j}$, jointly they can be written as 
\begin{equation}
    \tilde{\bm{\alpha}} = \tilde{\bm{B}}\tilde{\bm{\alpha}} + \tilde{\bm{\epsilon}},
    \label{eqn:2.2.8}
\end{equation}
where $\tilde{\bm{\alpha}}$ $ = (\tilde{\bm{\alpha}}^{\top}_{1}, \cdots, \tilde{\bm{\alpha}}^{\top}_{p}, \bar{\bm{\alpha}}_{1}^{\top}, \dots, \bar{\bm{\alpha}}_{p}^{\top})^{\top}$, $\tilde{\bm{\epsilon}} = (\tilde{\bm{\epsilon}}^{\top}_{1}, \cdots, \tilde{\bm{\epsilon}}^{\top}_{p}, \bm{\gamma}_{1}^{\top}, \dots, \bm{\gamma}_{p}^{\top})^{\top}$ with $ \tilde{\bm{\epsilon}}_{j} = (\tilde{\epsilon}_{j1}, \cdots, \tilde{\epsilon}_{jK_{j}})^{\top}$ and $\bm{\gamma}_{j} = (\gamma_{j,K_{j} + 1}, \dots, \gamma_{jS})^{\top}$. Here $\tilde{\bm{B}} = \begin{pmatrix}
    \bm{B} & \bm{0}\\
    \bm{0} & \bm{0}
\end{pmatrix}$ where $\bm{B} = ((\bm{B}_{j\ell}(a, b))_{a=1, b=1}^{K_{j}, K_{l}})_{j, \ell = 1}^{p}$ with $\bm{B}_{jj} = \bm{0}$ since we assume the absence of self loops. To carry out inference, we assume $\tilde{\epsilon}_{jk}, \gamma_{jk} \overset{ind}\sim \sum_{m = 1}^{M_{jk}} \pi_{jkm} {N}(\mu_{jkm}, \tau_{jkm})$.

\subsection{Adaptive basis expansion}

As the $\phi_{k}$'s are specifically useful for restricting the original function space for each $Y_{j}$ to a lower-dimensional causally embedded smooth space of dimension $K_{j}$, we make the basis $\{\phi_{k}\}$ adaptive for causal structure learning by further expanding them with known spline basis functions \citep{kowal2017bayesian}, $\phi_{k}(\cdot) = \sum_{r=1}^{R} A_{kr}b_{r}(\cdot)$, where $\bm{b} = (b_1, \dots , b_R)^{\top}$ is the set of fixed cubic B-spline basis functions with equally spaced knots and $\bm{A}_k = (A_{k1},\dots,A_{kR})^{\top}$ are the corresponding spline coefficients. Since we do not fix $\bm{A}_k$'s \emph{a priori}, the basis functions $\phi_k$'s can be learned from data \emph{a posteriori} and hence are adaptive to both data and causal structure (i.e., the basis functions, the functional data, and the causal graph are dependent in their joint distribution). 

\subsection{Prior specifications}

\paragraph{Prior on spline coefficients.}
The prior on $A_{k}$ is chosen to serve multiple purposes. (i) It sorts the basis functions by decreasing smoothness and therefore helps to identify the spanning set of size $K_{j}$ for the underlying smooth causally embedded space $\mathcal{D}_{j}$. 
(ii) Although not a strict requirement for modelling purpose, it forces $\phi_{k}$'s to be orthonormal, i.e. $\int \phi_{k}(\omega)\phi_{k'}(\omega)\,d\omega = I(k = k')$. As such, the orthogonality constraints help eliminate any information overlap between the basis functions, which keeps the total number of necessary basis functions that actually contribute to the causal structure learning to a minimum. (iii) It regularizes the roughness of $\phi_{k}$’s to prevent overfitting. 

For (iii), more specifically, we restrict the roughness of the basis functions $\phi_k(\cdot)$ by assigning a prior that penalizes its second derivatives \citep{gu1992penalized, wahba1978, berry2002splines}: 
   \begin{equation*}
       \bm{A}_{k} \sim N(\bm{0}, \lambda^{-1}_{k}\bm{\Omega^{-}}),
   \end{equation*}
   where $\bm{\Omega^{-}}$ is the pseudoinverse of $\bm{\Omega} = \int \bm{b}^{''}(t)[\bm{b}^{''}(t)]^{\top}\, dt$.
   Let $\bm{\Omega = UDU}^{\top}$ be the singular value decomposition of $\bm{\Omega}$. Following \citealp{wand}, to facilitate computation, we reparameterize $  \phi_{k}(\cdot) = \sum\limits_{r=1}^{R} \tilde{A}_{kr}\tilde{b}_{k}(\cdot)$ with $\tilde{\bm{b}}(\cdot) = (1, t, \bm{b}^{T}(\cdot)\tilde{\bm{U}}\tilde{\bm{D}}^{-\frac{1}{2}})^{\top}$ where $\tilde{\bm{D}}$ is the $(R - 2) \times (R - 2)$ submatrix of $\bm{D}$ corresponding to non-zero singular values (note that the rank of $\bm{\Omega}$ is $R-2$ by definition) and $\tilde{\bm{U}}$ is the corresponding $R \times (R - 2)$ submatrix of $\bm{U}$. This induces a prior on $\tilde{\bm{A}}_{k}$ given by
   \begin{equation*}
       \tilde{\bm{A}}_{k} \sim N(\bm{0}, \bm{S}_{k}) ~\text{with}~ \bm{S_{k}} = \text{diag}(\infty, \infty, \lambda^{-1}_{k}, \dots, \lambda^{-1}_{k}).
   \end{equation*}
In other words, the intercept and the linear term are unpenalized but the non-linear terms are penalized, the degree of which is controlled by $\lambda_k$.
In practice, we set the first two diagonal elements of $\bm{S}_{k}$ as $10^{8}$. We constrain the regularization parameters $\lambda_{1} > \dots > \lambda_{S} > 0$ by putting a uniform prior:
   \begin{equation*}
       \begin{aligned}
          \lambda_{k} &\sim \text{Uniform}(L_{k}, U_{k}), ~\forall~k \in [S],\\
          U_{1} &= 10^{8}, L_{k} = \lambda_{k+1} ~\forall~ k \in [S-1],\\
          U_{k} &= \lambda_{k-1} ~\forall~ k \in \{2, \dots, S\}, L_{S} = 10^{-8},
       \end{aligned}
   \end{equation*}
   which implies that the smoothness of $\phi_k(\cdot)$ decreases as $k$ gets larger.

\paragraph{Priors on the adjacency matrix.} We propose to use an independent uniform-Bernoulli prior on each entry $E_{j\ell}$ of $\bm{E}$, i.e., $E_{j\ell}|\rho \overset{\text{ind}}{\sim} \text{Bernoulli}(\rho)$ and $\rho \sim \text{Uniform}(0,1)$.
The marginal distribution of $\bm{E}$ with $\rho$ integrated out is given by 
\begin{equation*}
    p(\bm{E}) = \int p(\bm{E}|\rho)p(\rho)\, d\rho = \text{Beta}\left(\sum\limits_{j\neq \ell}E_{j\ell} + 1, \sum\limits_{j\neq \ell}(1- E_{j\ell}) + 1\right). 
\end{equation*}
Now, for example, if $\bm{E}_{0}$ denotes the null adjacency matrix and $\bm{E}_{1}$ denotes the adjacency matrix with only one edge, then we can see that $p(\bm{E}_{0})/p(\bm{E}_{1}) = p^{2} - p$. Therefore, an empty graph is favored over a graph with one edge by a factor of $p^{2} - p$, and, importantly, this penalty increases with $p$. Thus, the uniform-Bernoulli prior prevents false discoveries and leads to a sparse network by increasing the penalty against
additional edges as the dimension $p$ grows.  

\paragraph{Prior on the causal effect matrix.} Now given $\bm{E}$, we assume an independent spike and slab prior on the entries of $\bm{B} = (\bm{B}_{j\ell})_{j,\ell = 1}^{p}$:
\begin{equation*}
    \bm{B}_{j\ell}|E_{j\ell} \sim (1 - E_{j\ell})MVN(\bm{B}_{j\ell}; \bm{0}, s\gamma \bm{I}_{K_{j}}, \bm{I}_{K_{\ell}}) + E_{j\ell} MVN(\bm{B}_{j\ell}; \bm{0}, \gamma \bm{I}_{K_{j}},\bm{I}_{K_{\ell}}),
\end{equation*}
where $MVN(\bm{B}_{j\ell}; \bm{0}, \gamma \bm{I}_{K_{j}},\bm{I}_{K_{\ell}})$ is a matrix-variate normal distribution with row and column covariance matrices as $\gamma\bm{I}_{K_{j}}$ and $\bm{I}_{K_{\ell}}$, respectively. We assume a conjugate inverse-gamma prior on the causal effect size, $\gamma \sim \text{InverseGamma}(a_{\gamma},b_{\gamma})$. We choose $a_{\gamma} = b_{\gamma} = 1$. We fix $s = 0.02$ so that when $E_{j\ell}=0$, $\bm{B}_{j\ell}$ is negligibly small. 
 
\paragraph{Priors on the parameters of the Gaussian mixture distribution.} 
   We choose conjugate priors for the parameters of the Gaussian mixture distribution:
   \begin{equation*}
       \begin{aligned}
          &(\pi_{jk1}, \dots, \pi_{jkM_{jk}}) \sim\text{Dirichlet}(\beta, \dots, \beta), ~~~~~\forall ~ j \in [p], k \in [S]\\
          &\mu_{jkm} \sim N(a_{\mu}, b_{\mu}), ~\tau_{jkm} \sim \text{InverseGamma}(a_{\tau}, b_{\tau}), ~\forall ~ j \in [p], k \in [S], m \in [M_{jk}]
       \end{aligned}
   \end{equation*}
We have fixed values for the hyperparameters, $\beta = 1, a_{\mu} = 0, b_{\mu} = 100, a_{\tau} = b_{\tau} = 1$.
 
\paragraph{Prior on the noise variances.} We assume a conjugate prior for $\sigma_{j}\sim \text{InverseGamma}(a_{\sigma}, b_{\sigma}),\\ ~\forall~ j \in [p]$. We choose $a_{\sigma} = b_{\sigma} = 0.01$.

We simulate posterior samples through Markov chain Monte Carlo (MCMC). Details are given in Section \ref{posterior} of the Supplementary Materials. Sensitivity analyses will be conducted to test the hyperparameters including $(a_{\gamma}, b_{\gamma}), (a_{\tau}, b_{\tau}), (a_{\sigma}, b_{\sigma}), s, R, S, M$ and $\beta$.

\section{Simulation Studies} 
   \label{simulation_study}
\paragraph{Data generation} The data were simulated according to various combinations of sample size ($n$), number of nodes ($p$), and grid size
($m_{j} = d ~\forall j\in [p]$) where $n \in \{75, 150, 300\}$, $p \in \{20, 40, 60\}$, and $d \in \{125, 250\}$. The grid evenly spans the unit interval $[0, 1]$; the results with unevenly spaced grids are presented in Section \ref{additionalsimulations} of the Supplementary Materials. The true causal graph $\mathcal{G}$ was generated randomly with edge formation probability $2/p$. Given $\mathcal{G}$, each non-zero block $\bm{B}_{j\ell}$ of the causal effect matrix was generated from the standard matrix-variate normal distribution. We set the true number of basis functions to be $K = 4$. In order to generate $K = 4$ orthonormal basis functions, we first simulated unnormalized basis functions by expanding them further with 6 cubic B-spline basis functions where the coefficients were drawn from the standard normal distribution and then empirically orthonormalized them.  
The basis coefficients $\tilde{\bm{\alpha}}$ were generated following (\ref{eqn:2.2.8}) with the exogenous variables $\tilde{\bm{\epsilon}}_{j}$ drawn independently from Laplace distribution with location parameter $\mu = 0$ and scale parameter $b = 0.2$. We have also considered other non-Gaussian distributions for the exogenous variables; the corresponding results are provided in Section \ref{additionalsimulations} of the Supplementary Materials. Finally, noisy observations were simulated following (\ref{eqn:2.2.6}) with the signal-to-noise ratio, i.e., the mean value of $|Y^{(i)}_{j}(t)| / \sigma_{j}$ across all $i$ and $t$, set to 5. Here, superscript $(i)$ denotes the $i$th sample, where $i\in[n]$. 

For the implementation of the proposed FENCE, we fixed the number of mixture components to be 10 and ran MCMC for 5,000 iterations (discarding the first 2,000 iterations as burn-in and retaining every 5th iteration after burn-in). The
causal graph $G$ was then estimated by using the median probability model \citep{mrule}, i.e., by thresholding the posterior probability of inclusion at 0.5.

\paragraph{Methods for comparison.} We  compared our method with fLiNG \citep{fBN}, a recently proposed directed acyclic graph (DAG) for multivariate functional data. Codes for \cite{fSEM, pmlr-v186-yang22a} are not publicly available. Hence for more comparison, we considered two \emph{ad hoc} two-step  approaches. In the first step of both approaches, we obtained the basis coefficients by carrying out functional principal component analysis (fPCA) using the \texttt{fdapace} \citep{fdapace} package in R. Then in the second step, given the basis coefficients, we estimated causal graphs using  existing causal discovery methods, (i) LiNGAM \citep{shimizu06a} (ii) PC \citep{PCalgo} and (iii) CCD \citep{ccd};  we call these three approaches fPCA-LiNGAM, fPCA-PC and fPCA-CCD respectively. Note that we did not use SEM-based cyclic discovery algorithm, LiNG-D \citep{Lacerda} in the second step due to the unavailability of the code. 

LiNGAM estimates a causal DAG based on the linear non-Gaussian assumption whereas PC generally returns only an equivalence class of DAGs based on conditional independence tests. CCD algorithm is a constraint-based causal discovery method, which yields an equivalence class of cyclic causal graphs. LiNGAM and PC are implemented in the \texttt{pcalg} package \citep{Kalisch2018} in R. CCD algorithm is implemented in the \texttt{py-tetrad} \citep{ramsey2018tetrad} package in python. 

\paragraph{Performance metrics.}To assess the graph recovery performance, we calculated the true positive rate (TPR), false discovery rate (FDR), and Matthew's correlation coefficient (MCC). For TPR and MCC, higher is better, whereas lower FDR is better.

\paragraph{Results.} Table \ref{tab:example} summarizes the results of 50 repeat simulations, demonstrating that the proposed FENCE model outperforms all competitors (fLiNG, fPCA-LiNGAM, and fPCA-CCD) across all combinations of $n$, $p$, and $d$. We provide the results of fPCA-PC in Section \ref{fulltable} of the Supplementary Materials, which are similar to those of fPCA-LiNGAM. We favored fPCA-PC and fPCA-CCD by counting a non-invariant edge between two nodes as a true positive as long as the two nodes are adjacent in the true graph. The superiority of FENCE is not unexpected for three reasons. First, fLiNG, fPCA-LiNGAM, and fPCA-PC are not specifically designed for learning cyclic graphs. Second,  two-step approaches like fPCA-LiNGAM, fPCA-PC, and fPCA-CCD do not necessarily capture the causally embedded space through the functional principal components. Third, although fPCA-CCD can handle cyclic graphs, it, being a two-step approach, fails to capture true functional dependencies. Overall, these findings provide strong evidence of the effectiveness of FENCE compared to existing methods.

\paragraph{Additional simulations.} We considered additional simulation  scenarios with unevenly spaced grids, general exogenous variable distributions, true acyclic graphs and data generated using non-linear SEM, and also conducted sensitivity analyses of FENCE with respect to several hyperparameters; the results are presented in Section \ref{additionalsimulations} of the Supplementary Materials.
The performance of the proposed method is consistently better than competing methods and is relatively robust with respect to hyperparameter choices.

\begin{table}[htb]
\caption{Comparison of performance of various methods under 50 replicates. Since LiNGAM is not applicable to cases where $q > n$ with $q = Kp$ being the total number of extracted basis coefficients across all functions, the results from
those cases are not available and indicated by "-". The metrics reported are based on 50 repetitions are reported; standard deviations are given within the
parentheses.}
  \label{tab:example}
  \resizebox{\columnwidth}{!}{\begin{tabular}{ccc|*3c|*3c|*3c|*3c}
    \toprule
    \multirow{2}{*}{n}& \multirow{2}{*}{p} & \multirow{2}{*}{d} & \multicolumn{3}{c|}{FENCE}  & \multicolumn{3}{c|}{fLiNG} & \multicolumn{3}{c|}{fPCA-LINGAM} & \multicolumn{3}{c}{fPCA-CCD}\\  \cmidrule(l){4-6} \cmidrule(l){7-9} \cmidrule(l){10-12} \cmidrule(l){13-15} 
     &  &  & TPR & FDR & MCC & TPR & FDR & MCC & TPR & FDR & MCC & TPR & FDR & MCC \\ \hline
    75 & 20 & 125 & \textbf{0.85(0.09)} & \textbf{0.19(0.07)} & \textbf{0.88(0.05)} & 0.41(0.09) & 0.79(0.05) & 0.36(0.04) & 0.35(0.19) & 0.84(0.04) & 0.11(0.08) &  0.69(0.03) & 0.41(0.04) & 0.23(0.03)\\
    75 & 40 & 125 & \textbf{0.79(0.08)} & \textbf{0.23(0.06)} & \textbf{0.86(0.04)} & 0.37(0.08) & 0.82(0.06) & 0.33(0.05) & - & - & - & 0.73(0.02) & 0.47(0.04)& 0.21(0.05)\\ 
    75 & 60 & 125 & \textbf{0.75(0.07)} & \textbf{0.27(0.05)} & \textbf{0.83(0.04)} & 0.34(0.07) & 0.83(0.06) & 0.32(0.04) &  - & - & - & 0.68(0.03) & 0.61(0.05) & 0.19(0.03)\\ \midrule
    150 & 20 & 125 & \textbf{0.88(0.07)} & \textbf{0.14(0.06)} & \textbf{0.89(0.05)} & 0.45(0.07) & 0.75(0.06) & 0.39(0.05) & 0.28(0.22) & 0.86(0.05) & 0.08(0.09) &  0.71(0.03) & 0.42(0.03) & 0.25(0.04)\\
    150 & 40 & 125 & \textbf{0.81(0.07)} & \textbf{0.21(0.06)} & \textbf{0.87(0.05)} & 0.39(0.06) & 0.79(0.05) & 0.37(0.04) & 0.35(0.22) & 0.91(0.02) & 0.08(0.06) &  0.73(0.04) & 0.47(0.05) & 0.23(0.03)\\
    150 & 60 & 125 & \textbf{0.79(0.06)} & \textbf{0.24(0.05)} & \textbf{0.86(0.04)} & 0.36(0.04) & 0.80(0.06) & 0.36(0.05)  & - & - & - & 0.72(0.05) & 0.54(0.04) & 0.22(0.02)\\ \midrule
    300 & 20 & 125 & \textbf{0.91(0.03)} & \textbf{0.09(0.04)} & \textbf{0.90(0.04)} & 0.51(0.04) & 0.73(0.06) & 0.41(0.04) & 0.30(0.19) & 0.84(0.05) & 0.11(0.09) & 0.81(0.03) & 0.39(0.04) & 0.26(0.03)\\
    300 & 40 & 125 & \textbf{0.87(0.04)} & \textbf{0.15(0.05)} & \textbf{0.87(0.05)} & 0.47(0.05) & 0.75(0.06) & 0.38(0.05) & 0.27(0.20) & 0.91(0.02) & 0.08(0.06) &  0.77(0.03) & 0.45(0.02) & 0.24(0.03)\\
    300 & 60 & 125 & \textbf{0.85(0.05)} & \textbf{0.17(0.03)} & \textbf{0.86(0.03)} & 0.45(0.05) & 0.76(0.04) & 0.38(0.03) &  0.28(0.17) & 0.91(0.05)  & 0.05(0.03) & 0.72(0.03) & 0.49(0.02) & 0.22(0.03)\\
    \midrule
    75 & 20 & 250 & \textbf{0.81(0.04)} & \textbf{0.23(0.02)} & \textbf{0.85(0.05)} & 0.39(0.07) & 0.80(0.05) & 0.39(0.04) & 0.32(0.14) & 0.82(0.03) & 0.09(0.04) & 0.67(0.03) & 0.46(0.03) & 0.22(0.04) \\
    75 & 40 & 250 & \textbf{0.73(0.04)} & \textbf{0.28(0.05)} & \textbf{0.82(0.04)} & 0.35(0.04) & 0.85(0.06) & 0.33(0.05) & - & - & - & 0.68(0.02) & 0.51(0.04) & 0.21(0.03)\\ 
    75 & 60 & 250 & \textbf{0.67(0.03)} & \textbf{0.34(0.05)} & \textbf{0.79(0.04)} & 0.34(0.04) & 0.85(0.03) & 0.31(0.04) &  - & - & -  & 0.63(0.04) & 0.56(0.04) & 0.19(0.04)\\ \midrule
    150 & 20 & 250 & \textbf{0.83(0.06)} & \textbf{0.17(0.05)} & \textbf{0.86(0.05)} & 0.46(0.07) & 0.73(0.07) & 0.42(0.05) & 0.32(0.19) & 0.79(0.05) & 0.13(0.05) &  0.73(0.04) & 0.43(0.02) & 0.24(0.03)\\
    150 & 40 & 250 & \textbf{0.79(0.02)} & \textbf{0.26(0.06)} & \textbf{0.82(0.03)} & 0.41(0.05) & 0.71(0.05) & 0.40(0.03) & 0.31(0.14) & 0.81(0.02) & 0.13(0.06) &  0.71(0.03) & 0.47(0.04) & 0.23(0.03)\\
    150 & 60 & 250 & \textbf{0.69(0.05)} & \textbf{0.31(0.05)} & \textbf{0.79(0.04)} & 0.43(0.03) & 0.79(0.06) & 0.43(0.05)  & - & - & - & 0.69(0.02) & 0.52(0.03) & 0.21(0.02)\\ \midrule
    300 & 20 & 250 & \textbf{0.86(0.02)} & \textbf{0.16(0.04)} & \textbf{0.85(0.04)} & 0.68(0.02) & 0.77(0.07) & 0.47(0.04) & 0.45(0.13) & 0.86(0.05) & 0.17(0.09) & 0.78(0.02) & 0.44(0.06) & 0.27(0.03)\\
    300 & 40 & 250 & \textbf{0.79(0.08)} & \textbf{0.16(0.05)} & \textbf{0.84(0.06)} & 0.73(0.05) & 0.71(0.06) & 0.43(0.05) & 0.39(0.16) & 0.87(0.05) & 0.16(0.07) & 0.76(0.05) & 0.49(0.06) & 0.23(0.05)\\
    300 & 60 & 250 & 0.76(0.05) & \textbf{0.21(0.03)} & \textbf{0.80(0.03)} & \textbf{0.77(0.05)} & 0.74(0.03) & 0.42(0.03) &  0.28(0.17) & 0.90(0.04) & 0.13(0.04) & 0.72(0.06) & 0.53(0.03) & 0.22(0.04)\\
    \bottomrule
  \end{tabular}}
 
\end{table}

\section{Real Data Application}
   \paragraph{Brain EEG data.} We demonstrate the proposed FENCE model on a brain EEG dataset from an alcoholism study \citep{Zhang1995EventRP}. This dataset was earlier used to demonstrate functional undirected graphical models \citep{zhu2016bayesian, qiao2019functional} and functional Bayesian network \citep{fBN}. Data were initially obtained from 64 electrodes placed on subjects’ scalps, which captured EEG signals at 256 Hz (3.9 ms epoch) during a one-second period. The study consists of 122 subjects, out of which 77 are in the alcoholic group and 45 are in the control group. Each subject completed 120 trials. During each trial, the subject was exposed to either a single stimulus (a single picture) or two stimuli (a pair of pictures) shown on a computer monitor. We particularly focus on the EEG signals filtered at $\alpha$ frequency bands between 8 and 12.5Hz using the \texttt{eegfilt} function of the \texttt{eeglab} toolbox of Matlab as $\alpha$ band signals are associated with inhibitory control \citep{knyazev2007motivation}. Given that the EEG measurements were recorded from each subject over multiple trials, these measurements are not independent of each other due to the time dependency of the trials. Moreover, since the measurements were obtained under various stimuli, the signals may have been affected by different stimulus effects. To mitigate these issues, we calculated the average of the band-filtered EEG signals for each subject across all trials under a single stimulus, resulting in a single event-related potential  curve per electrode per subject. By doing so, we eliminated the potential dependence between the measurements and the influence of different stimulus types. We performed separate analyses of the two groups to identify both the similarities and dissimilarities in their brain effective connectivity.

We conducted a Shapiro-Wilk normality test on the observed functions for each of the $p = 64$ scalp positions at each of the $m_{j} = 256~\forall j\in[p]$ time points to evaluate their Gaussianity. The results showed that for numerous combinations of scalp position and time point, the null hypothesis (which assumes that the observations are marginally Gaussian) was rejected. Thus, we conclude that the non-Gaussianity of the proposed model is appropriate. Next, for posterior inference, we ran MCMC for 20,000 iterations, discarded the first half as burn-in, and retained every 10th iteration after burn-in. The estimated causal graph by thresholding the posterior inclusion probability to $0.9$ is given below in Figure \ref{fig:2}. 

\begin{figure}[!ht]
    \centering
    \resizebox{\columnwidth}{!}{\includegraphics[width=10cm]{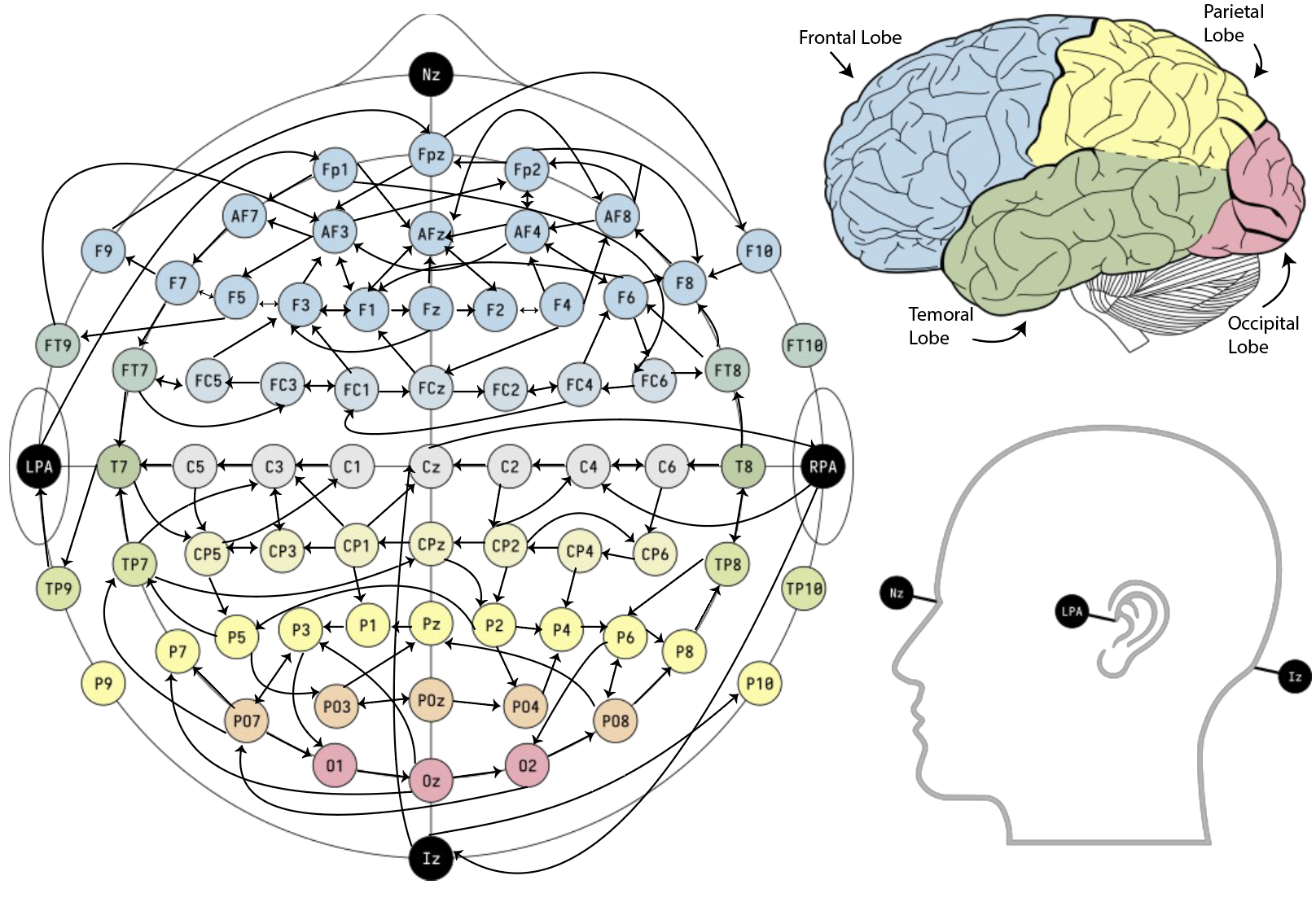}
    \includegraphics[width=6cm, height = 7cm]{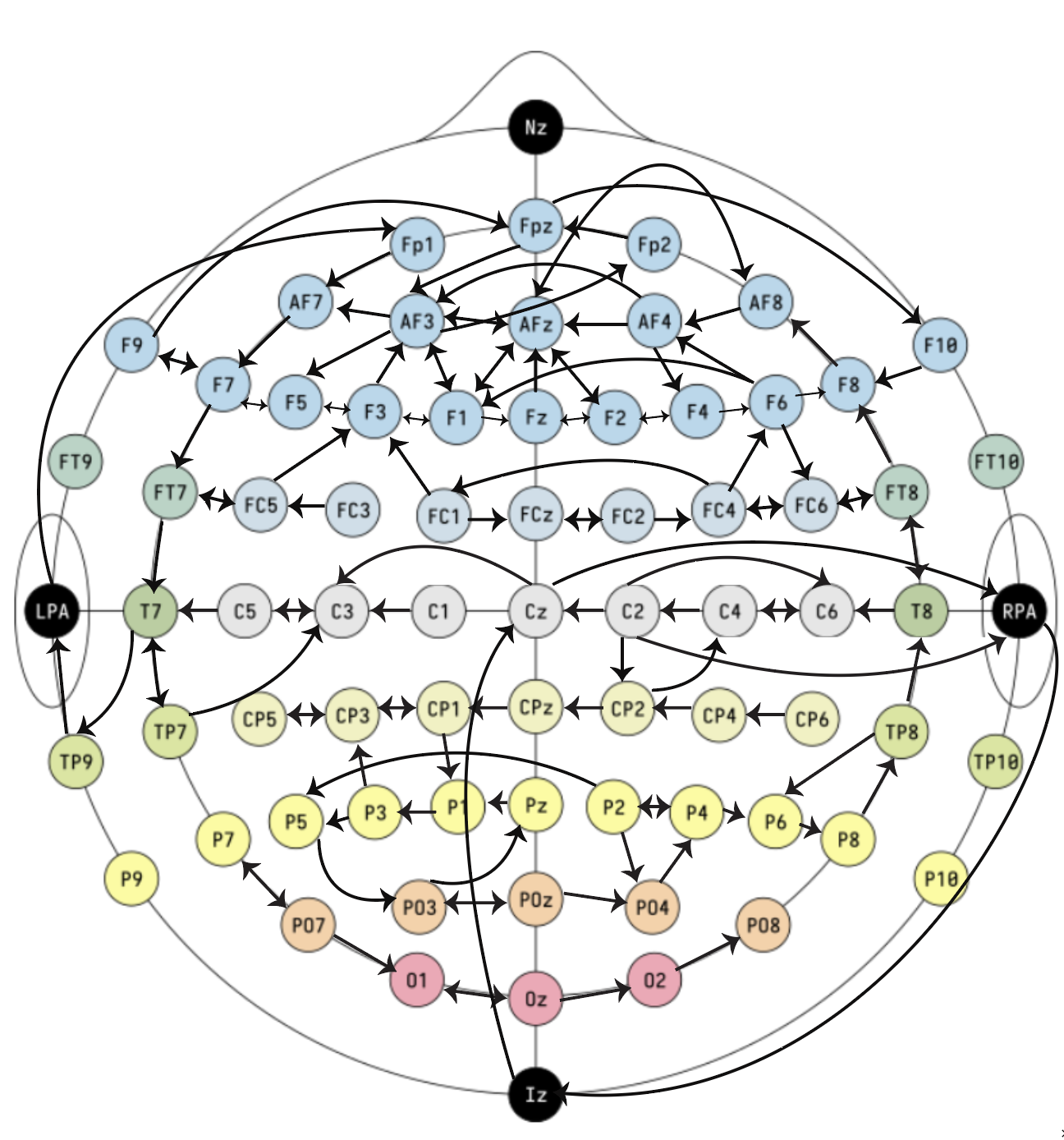}}
    \caption{Estimated causal brain connectivity from EEG records by FENCE with posterior probability of inclusion $\geq 0.9$, separately for the alcoholic (left) and control (right) group. The bi-directed edges are just directed cycles, i.e., $i \leftrightarrow j$ means $i \rightarrow j$  and $i \leftarrow j$.}
    \label{fig:2}
\end{figure}

\paragraph{Results.} There are some interesting findings. First, for both groups (alcoholic and control), brain regions that are located in adjacent positions tend to be more connected than the brain regions that are far apart. Second, dense connectivity is observed in the frontal region of the brain in both groups, with multiple cycles being formed. Third, compared to the control group, the alcoholic group has more connectivity across the left parietal and occipital lobes. Fourth, the same cycle of Iz, Cz, and RPA is observed in both groups. 

\paragraph{Validity.}  We now discuss the validity of our real data results. In \citealp{hayden}, it was observed that alcohol-dependent subjects exhibited frontal asymmetry, distinguishing them from the control group. Our own investigation aligns well with these results, as we have identified denser connectivity across various brain regions in the middle and left areas of the frontal lobe among alcoholic subjects, when compared to controls. Furthermore, \citealp{winterer} documented coherent differences between alcoholics and controls in the posterior hemispheres, specifically in the temporal, parietal, and occipital lobes. In accordance with their findings, our study provides additional support for this claim, as we have observed heightened activity with several cycles formed in those same regions within the alcoholic group when compared to the control group.

\section{Discussion}
   We briefly highlight here several potential avenues for future development of our current work. First, an intriguing and important direction would be to explore the relaxation of the causal sufficiency assumption in the model identifiability. Second, our current model is based on a linear non-Gaussian assumption over the exogenous variables, but a nonlinear model could be considered as an alternative. Lastly, an alternative approach to determining the effective number of basis functions that span the lower-dimensional causal embedded space would be to utilize the ordered shrinkage priors \citep{bhattacharya, legramanti} in order to adaptively eliminate redundant components, resulting in a more flexible methodology. 
\section{Acknowledgement}
   Ni's research was partially supported by NSF DMS-2112943 and NIH 1R01GM148974-01. 
\bibliography{Reference}

\newpage
   \begin{center}
    \Large \textbf{Supplementary Materials for ``Directed Cyclic Graph for Causal Discovery from Multivariate
Functional Data"}
\end{center}
\appendix

\section{Proof of theorem \ref{Th:1}} \label{section:proof}
\begin{proof}
For some basis $\{\phi_{jk}\}_{k=1}^{K_{j}}$ that spans the low dimensional causal embedded space $\mathcal{D}_{j}$, $\alpha_{j}$ in (\ref{eqn:2.2.5}) of the main manuscript can be further expanded by $$\alpha_{j}(t_{ju}) = \sum_{k = 1}^{K_{j}}\tilde{\alpha}_{jk}\phi_{jk}(t_{ju})$$ Using the above , (\ref{eqn:2.2.5}) can then be expressed as, 
\begin{align}
    X_{ju} &= \sum_{k = 1}^{K_{j}}\tilde{\alpha}_{jk}\phi_{jk}(t_{ju}) + \beta_{j}(t_{ju}) + e_{ju}, \forall j\in [p], u \in [m_{j}]
    \label{eqn:A.0.2}
\end{align}
More compactly, the above (\ref{eqn:A.0.2}) can be rewritten as,
\begin{align}
    \bm{X} &= \bm{\Phi}(\bm{t})\tilde{\bm{\alpha}} + \bm{\beta}(\bm{t}) + \bm{e} 
    \label{eqn:A.0.3}
\end{align}
where $\bm{X} = (\bm{X}^{\top}_{1}, \dots, \bm{X}^{\top}_{p})^{\top}, \tilde{\bm{\alpha}} = (\tilde{\bm{\alpha}}^{\top}_{1}, \dots, \tilde{\bm{\alpha}}^{\top}_{p})^{\top}, \bm{\beta}(\bm{t}) = (\bm{\beta}_{1}(\bm{t}_{1})^{\top}, \dots, \bm{\beta}_{p}(\bm{t}_{p})^{\top})^{\top}, \bm{e} = (\bm{e}^{\top}_{1}, \dots, \bm{e}^{\top}_{p})^{\top}$ and $\bm{\Phi}(\bm{t}) = \diag(\bm{\Phi}_{1}(\bm{t}_{1}), \dots, \bm{\Phi}_{p}(\bm{t}_{p}))$ with $\bm{X}_{j} = (X_{j1}, \dots, X_{jm_{j}})^{\top}, \tilde{\bm{\alpha}}_{j} = (\tilde{\alpha}_{j1}, \dots, \tilde{\alpha}_{jK_{j}})^{\top}, \bm{\beta}_{j}(\bm{t}_{j}) = (\beta_{j}(t_{j1}), \dots, \beta
_{j}(t_{jm_{j}}))^{\top}, \bm{e}_{j} = (e_{j1}, \dots, e_{jm_{j}})^{\top}$ and 
$$ \bm{\Phi}_{j}(\bm{t}_{j}) = \begin{pmatrix*}
\phi_{j1}(t_{j1}) & \phi_{j2}(t_{j1}) & \cdots & \phi_{jK_{j}}(t_{j1}) \\
\phi_{j1}(t_{j2}) & \phi_{j2}(t_{j2}) & \cdots & \phi_{jK_{j}}(t_{j2}) \\
\vdots & \vdots & \ddots & \vdots \\
\phi_{j1}(t_{jm_{j}}) & \phi_{j2}(t_{jm_{j}}) & \dots & \phi_{jK_{j}}(t_{jm_{j}}) \\
\end{pmatrix*}$$
The structural equation model is then defined on $\tilde{\bm{\alpha}}$ as, 
\begin{align}
   \tilde{\bm{\alpha}} &= \bm{B}\tilde{\bm{\alpha}} + \tilde{\bm{\epsilon}} \nonumber\\
   \Rightarrow \tilde{\bm{\alpha}} &= \bm{\Omega}\tilde{\bm{\epsilon}}, ~~ [\text{by Assumption \ref{ass3}}]
   \label{eqn:A.0.4}
\end{align}
where $\bm{\Omega} = \bm{(I - B)^{-1}}$. 

Referring to Assumption \ref{ass5} of section \ref{sec:2.3} in the main manuscript, we write $\bm{\beta}(\bm{t}) = \bm{C}(\bm{t})\bm{\gamma}$ where $\gamma_{jk} \sim \sum\limits_{m = 1}^{M_{jk}}\pi_{jkm}'N(\mu_{jkm}', \tau_{jkm}')$ with $$\bm{C} = \begin{pmatrix*}
 \bm{C}_{11}(\bm{t}_{1})& \bm{0}& \cdots & \bm{0}  \\
 \bm{0} & \bm{C}_{22}(\bm{t}_{2}) & \cdots & \bm{0}\\
 \vdots & \vdots & \ddots & \vdots \\
 \bm{0} & \bm{0} & \cdots & \bm{C}_{pp}(\bm{t}_{p})
\end{pmatrix*}$$ 
Using this representation for $\bm{\beta}(\bm{t})$ and (\ref{eqn:A.0.4}), (\ref{eqn:A.0.3}) boils down to, 
\begin{equation}
    \label{eqn:A.0.5}
     \bm{X} = \bm{\Phi}(\bm{t})\bm{\Omega}\tilde{\bm{\epsilon}} + \bm{C}(\bm{t})\bm{\gamma} + \bm{e} 
\end{equation}
From here on let us define $N = \sum_{j=1}^{p}m_{j}$ and $K = \sum_{j=1}^{p}K_{j}$. We define two class variables $\bm{\xi}$ and $\bm{\eta}$ such that $\epsilon_{jk}|\xi_{jk} = m \sim N(\mu_{jkm}, \tau_{jkm})$ and $\mathbb{P}(\xi_{jk} = m) = \pi_{jkm}$ and $\gamma_{jk}|\eta_{jk} = m \sim N(\mu_{jkm}', \tau_{jkm}')$ and $\mathbb{P}(\eta_{jk} = m) = \pi_{jkm}'$. Conditioning on these class variables $\bm{\xi}$ and $\bm{\eta}$,

\begin{align}
    \bm{X}|\bm{\xi}, \bm{\eta} &\sim N(\bm{\mu}_{\bm{X}}, \bm{\Sigma}_{\bm{X}})
\end{align}
where,
\begin{align*}
   \bm{\mu}_{\bm{X}} &= \bm{\Phi}(\bm{t})\bm{\Omega}\bm{\mu_{\xi}} + \bm{C}(\bm{t})\bm{\mu_{\eta}} \\
   \bm{\Sigma}_{\bm{X}} &= \bm{\Phi}(\bm{t})\bm{\Omega}\bm{T_{\xi}}\bm{\Omega}^{\top}\bm{\Phi}(\bm{t})^{\top} + \bm{C}(\bm{t})\bm{T_{\eta}}\bm{C}(\bm{t})^{\top} + \bm{\Sigma}
\end{align*}

\noindent with $\bm{\mu_{\xi}} = (\bm{\mu_{\xi_{1}}}^{\top}, \dots, \bm{\mu_{\xi_{p}}}^{\top})^{\top}$ and $\bm{\mu}_{\bm{\eta}} = (\bm{\mu_{\eta_{1}}}^{\top}, \dots, \bm{\mu_{\eta_{p}}}^{\top})^{\top}$ are the collection of means and $\bm{T_{\xi}} = \diag(\bm{T_{\xi_{1}}}, \dots, \bm{T_{\xi_{p}}})$ and $\bm{T_{\eta}} = \diag(\bm{T_{\xi_{1}}}, \dots, \bm{T_{\xi_{p}}})$ are diagonal matrices with variances as diagonal entries corresponding to the class variable $\bm{\xi}$ and $\bm{\eta}$. Here, $\bm{\mu_{\xi_{j}}} = (\mu_{\xi_{j1}}, \dots, \mu_{\xi_{jK_{j}}})^{\top}, \bm{\mu_{\eta_{j}}} = (\mu_{\eta_{j1}}, \dots, \mu_{\eta_{jK_{j}}})^{\top}, \bm{T_{\xi_{j}}} = \diag(T_{\xi_{j1}}, \dots, T_{\xi_{jK_{j}}})$ and $\bm{T_{\eta_{j}}} = \diag(T_{\eta_{j1}}, \dots, T_{\eta_{ju}}, \dots)$ with $\mu_{\xi_{jk}} = \mu_{jkm}$ if $\xi_{jk} = m$ and $T_{\xi_{jk}} = \tau_{jkm}$ if $\xi_{jk} = m$ and $\mu_{\eta_{jk}} = \mu_{jkm}'$ if $\eta_{jk} = m$ and $T_{\eta_{jk}} = \tau_{jkm}'$ if $\eta_{jk} = m$. $\bm{\Sigma}^{N \times N} = \diag(\sigma_{1}, \dots, \sigma_{1}, \dots, \sigma_{p}, \dots, \sigma_{p})$. 

Our causal identifiability proof necessarily involves \textbf{two steps} - First, we shall prove that the hypergraph like structure which is formed under the assumption of existence of disjoint cycles (Refer Assumption \ref{ass2} of the main manuscript) is identifiable. Second, the disjoint cycles inside every hypernode are identifiable. Please refer to Figure \ref{fig:1} in the main paper for an artistic exposition of the proof structure. 

\paragraph{Step 1.} Now we shall prove the identifiability of our model under the assumption that the SEM involving $\tilde{\bm{\alpha}}$ has an underlying graph in which the cycles are disjoint (Assumption \ref{ass2}). Under this assumption, we will have cycles of variable length which are connected by directed edges such that no two cycles in the graph have two nodes that are common to both. This induces a hypergraph like structure with each disjoint cycle forming a simple directed cycle in $\mathcal{V}$. 

 Let us mathematically formalize what we have discussed in the above paragraph. Suppose, $\mathfrak{C} = \{\mathcal{C}_{1}, \dots, \mathcal{C}_{u}\}$ where each $\mathcal{C}_{i}$ is a simple directed cycle. Clearly, $\mathcal{V} = \cup_{i=1}^{u}\mathcal{C}_{i}$ as $\mathcal{C}_{i}$s form a partition in $\mathcal{V}$. Without loss of generality, let us assume that $\{\tilde{\bm{\alpha}}_{1}, \dots, \tilde{\bm{\alpha}}_{p}\}$ be arranged in such a way that the first $r_{1}$ elements form the simple cycle $\mathcal{C}_{1}$, the next $r_{2}$ elements form another simple cycle $\mathcal{C}_{2}$ and so on such that $\sum_{i=0}^{u}r_{i} = p$ with $r_{0} = 0$ and $\mathcal{C}_{i} = \{\tilde{\bm{\alpha}}_{r_{i-1} + 1}, \dots, \tilde{\bm{\alpha}}_{r_{i}}\}$. We denote the hypergraph formed by $\mathfrak{C}$ by $\bar{\mathcal{G}}$.

Let $\bar{\mathcal{G}}$ and $\bar{\mathcal{G'}}$ be two graphs where $\bar{\mathcal{G'}} \neq \bar{\mathcal{G}}$. We can assume a topological ordering in $\bar{\mathcal{G}}$ in a sense that if $\bm{\mathcal{C}_{q}} \rightarrow \bm{\mathcal{C}_{r}}$ then $q<r$. Therefore, the $\bm{B}$ induced by the graph $\bar{\mathcal{G}}$ is necessarily a lower block triangular matrix with block $\bm{0}$ as the diagonal entries. We cannot say any such thing about the matrix $\bm{B'}$ induced by the graph $\bar{\mathcal{G'}}$ except that having block $\bm{0}$ matrices as it's diagonal elements. 

Let $\mathbb{P}$ and $\mathbb{P'}$ be the joint probability distribution of $\bm{X}$ associated with the two graphs $\mathcal{G}$ and $\mathcal{G'}$ respectively. Let $\mathcal{S} = (\bar{\mathcal{G}}, \mathbb{P})$ and $\mathcal{S'} = (\bar{\mathcal{G'}}, \mathbb{P'})$. We shall prove by contradiction that $\mathcal{S}$ and $\mathcal{S'}$ are not equivalent. 

Suppose, $\mathbb{P}(\bm{X}) \equiv \mathbb{P}'(\bm{X})$. Then due to the identifiability of finite Gaussian mixture models up to label permutation \citep{finiteMixtures1, finiteMixtures}, we must have, for any $\bm{\xi}, \bm{\eta}$,
\begin{align}
 \bm{\Phi}(\bm{t})\bm{\Omega}\bm{T_{\xi}}\bm{\Omega}^{\top}\bm{\Phi}(\bm{t})^{\top} + \bm{C}(\bm{t})\bm{T_{\eta}}\bm{C}(\bm{t})^{\top} + \bm{\Sigma} = \bm{\Phi}(\bm{t})\bm{\Omega}'\bm{T_{\xi}}'\bm{\Omega}'^{\top}\bm{\Phi}(\bm{t})^{\top} + \bm{C}(\bm{t})\bm{T_{\eta}}'\bm{C}(\bm{t})^{\top} + \bm{\Sigma}'
 \label{eqn:A.0.7}  
\end{align}

For some choice of $\tilde{\bm{\xi}} \neq \bm{\xi}$ and $\tilde{\bm{\eta}} = \bm{\eta}$, we can write from (\ref{eqn:A.0.7}), 
\begin{align}
\bm{\Phi}(\bm{t})\bm{\Omega}(\bm{T_{\xi}} - \bm{T_{\tilde{\xi}}})\bm{\Omega}^{\top}\bm{\Phi}(\bm{t})^{\top} &= \bm{\Phi}(\bm{t})\bm{\Omega}'(\bm{T_{\xi}}' - \bm{T_{\tilde{\xi}}}')\bm{\Omega}'^{\top}\bm{\Phi}(\bm{t})^{\top}\nonumber\\
\Rightarrow \bm{\Omega}(\bm{T_{\xi}} - \bm{T_{\tilde{\xi}}})\bm{\Omega}^{\top} &= \bm{\Omega}'(\bm{T_{\xi}}' - \bm{T_{\tilde{\xi}}}')\bm{\Omega}'^{\top}, (\text{using Assumption}~ \ref{ass6}) 
\label{eqn:A.0.8}
\end{align}

Notice that $\bm{\Omega}$ being an invertible matrix, every row of every block diagonal matrices must have at least a non zero element. $\bm{\Omega}_{K.}$  denotes the last row for $\bm{\Omega}$ and $l_{1}$ be the extreme position for which $\bm{\Omega}_{K, l_{1}} \neq 0$. Pick $\tilde{\bm{\xi}}$ above such that $\tilde{\bm{\xi}} = \bm{\xi}$ except for that $l_{1}$th element
such that $(\bm{T_{\xi}} - \bm{T_{\tilde{\xi}}})_{l_{1}, l_{1}} \neq 0$. Hence the matrix $(\bm{T_{\xi}} - \bm{T_{\tilde{\xi}}})$ is of rank 1 and from (\ref{eqn:A.0.8}) it implies that $\exists~ s_{1} \in [K]$ such that $(\bm{T}'_{\bm{\xi}} - \bm{T_{\tilde{\xi}}}')_{s_{1},s_{1}} \neq 0$. Therefore clearly,
\begin{eqnarray}
    0 \neq \bm{\Omega}_{K,.} \bm{(T_{\xi} - T_{\tilde{\xi}})}  \bm{\Omega}^{T}_{K,.} = \bm{\Omega}'_{K,.} (\bm{T}'_{\bm{\xi}} - \bm{T_{\tilde{\xi}}}')  \bm{\Omega}'^{\top}_{K,.} = \bm{\Omega}'^{2}_{K,s_{1}}(\bm{T}'_{\bm{\xi}} - \bm{T_{\tilde{\xi}}}')_{s_{1},s_{1}}
    \label{eqn:A.0.9}
\end{eqnarray}

Now as $(\bm{T}'_{\bm{\xi}} - \bm{T_{\tilde{\xi}}}')_{s_{1},s_{1}} \neq 0$, we have from (\ref{eqn:A.0.9}), $\bm{\Omega}'_{K,s_{1}} \neq 0$. Similarly, if we now focus on the $(K - 1)$th row of $\bm{\Omega}$, there can be two cases, 

\underline{\bf{Case 1:}} The last position for which $\bm{\Omega}_{K-1, .} \neq 0$ coincides with $l_{1}$. Then for this, we shall proceed with the same choice of $\tilde{\bm{\xi}}$ as above and with the same argument from above we can show that $\bm{\Omega}'_{K-1,s_{1}} \neq 0$ 

\underline{\bf{Case 2:}} If the position of the last non zero element in the $(K-1)$th row of $\bm{\Omega}$ is some $l_{2} (\neq l_{1})$, we pick $\tilde{\bm{\xi}}$ such that $\tilde{\bm{\xi}} = \bm{\xi}$ except for that $l_{2}$th element
such that $(\bm{T_{\xi}} - \bm{T_{\tilde{\xi}}})_{l_{2}, l_{2}} \neq 0$. Hence the matrix $(\bm{T_{\xi}} - \bm{T_{\tilde{\xi}}})$ is of rank 1 and from (\ref{eqn:A.0.8}) it implies that $\exists~ s_{2} \in [K]$ such that $(\bm{T'_{\xi}} - \bm{T^{'}_{\tilde{\xi}}})_{s_{2}, s_{2}} \neq 0$. Therefore clearly,
\begin{eqnarray}
    0 \neq \bm{\Omega}_{K-1,.} \bm{(T_{\xi} - T_{\tilde{\xi}})}  \bm{\Omega}^{T}_{K-1,.} = \bm{\Omega}'_{K-1,.} (\bm{T}'_{\bm{\xi}} - \bm{T_{\tilde{\xi}}}')  \bm{\Omega}'^{\top}_{K-1,.} = \bm{\Omega}'^{2}_{K-1,s_{2}}(\bm{T}'_{\bm{\xi}} - \bm{T_{\tilde{\xi}}}')_{s_{2},s_{2}}
    \label{eqn:A.0.10}
\end{eqnarray}

 Similarly as before since $(\bm{T'_{\xi}} - \bm{T^{'}_{\tilde{\xi}}})_{s_{2}, s_{2}} \neq 0$, we have from (\ref{eqn:A.0.10}), $\bm{\Omega}'_{K-1,s_{2}} \neq 0$. Define $K_{|\mathcal{C}_{i}|} = \sum_{j = r_{i-1}+1}^{r_{i}} K_{j}$. Clearly, $\sum_{i=1}^{u}\sum_{j = r_{i-1}+1}^{r_{i}} K_{j} = K$. Therefore, proceeding similarly from above we can show that $\bm{\Omega}'_{K-K_{|\mathcal{C}_{u}|}+1, s_{K_{|\mathcal{C}_{u}|}}} \neq 0$.  
 
 Now since $\bm{\Omega}$ is a lower block triangular matrix, we have $\forall ~ r \leq K - K_{|\mathcal{C}_{u}|}, \bm{\Omega}_{r, (K - K_{|\mathcal{C}_{u}|} + 1):K} = 0$. Therefore, if we pick some $\tilde{\bm{\xi}}$ which does not match $\bm{\xi}$ at the $l_{j}$th position, $l_{j} > K - K_{|\mathcal{C}_{u}|}, j \in [K_{|\mathcal{C}_{u}|}]$ such that $(\bm{T}_{\xi} - \bm{T}_{\tilde{\xi}})_{l_{j}, l_{j}} \neq 0$ then there will exist some $s_{j} \in [K]$ such that $(\bm{T'_{\xi}} - \bm{T^{'}_{\tilde{\xi}}})_{s_{j}, s_{j}} \neq 0, j \in [K_{|\mathcal{C}_{u}|}]$. Therefore we have,
\begin{eqnarray}
    0 = \bm{\Omega}_{r,.} \bm{(T_{\xi} - T_{\tilde{\xi}})}  \bm{\Omega}^{T}_{r,.} = \bm{\Omega}'_{r,.} (\bm{T}'_{\bm{\xi}} - \bm{T_{\tilde{\xi}}}')  \bm{\Omega}'^{\top}_{r,.} = \bm{\Omega}'^{2}_{r,s_{j}}(\bm{T}'_{\bm{\xi}} - \bm{T_{\tilde{\xi}}}')_{s_{j},s_{j}}
    \label{eqn:A.0.11}
\end{eqnarray}
 
From (\ref{eqn:A.0.11}), as $(\bm{T}'_{\bm{\xi}} - \bm{T_{\tilde{\xi}}}')_{s_{j},s_{j}} \neq 0$, we have, $\bm{\Omega}'_{r,s_{j}} = 0, \forall ~ r \leq K - K_{|\mathcal{C}_{u}|}, j \in [K_{|\mathcal{C}_{u}|}]$. 

Proceeding similarly from above, if we repeat the above set of arguments for all the rows of $\bm{\Omega}$ matrix, we can observe that $\bm{\Omega}'$ is just a block column permutation of a lower block triangular matrix. Therefore there exists a block lower triangular matrix $\bm{A}$ and a block permutation matrix $\bm{P}$ such that,
\begin{align}
  \bm{\Omega}' &= \bm{A P}\nonumber\\
  \Rightarrow \bm{(I - B')}^{-1} &= \bm{AP} \nonumber\\
  \Rightarrow \bm{(I - B')} &= \bm{P}^{\top} \bm{A}^{-1} 
  \label{eqn:A.0.12}
\end{align}

Now, the RHS of (\ref{eqn:A.0.12}) is just a row permuted block lower triangular matrix. Therefore, the permutation matrix $\bm{P}$ has to be the identity matrix; otherwise $\bm{P^{T} A^{-1}}$ must have zeros in its diagonal but $\bm{I}-\bm{B}'$ has unit diagonal because $\bm{B}'$ has zero diagonal (no self-loop). Hence we arrive at a contradiction and conclude from here that $\mathcal{S}$ and $\mathcal{S'}$ are not equivalent, i.e. $\mathbb{P}(\bm{X}) \neq \mathbb{P}'(\bm{X})$. 

\paragraph{Step 2.}We now try to prove that each simple directed cycle is identifiable.

If $\bm{H}$ and $\bm{H'}$ are the sub-matrices induced by some $\mathcal{C}_{j} \in \mathfrak{C}, j\in[u]$ in $\bar{\mathcal{G}}$  and $\bar{\mathcal{G'}}$ respectively then it is sufficient to show that for any permutation matrix $\bm{P}$,
$\bm{P(I - H)} = \bm{I} - \bm{H}' \Rightarrow \bm{P} = \bm{I}$.

Now since $\bm{H}$ is a matrix for a simple cycle, it can be written as $\bm{H = QD}$ where $\bm{Q}$ is a permutation matrix and $\bm{D}$ is block diagonal matrix. Now,

\begin{align}
    \bm{P}(\bm{I} - \bm{H}) &= \bm{I} - \bm{H}'\nonumber\\
\Rightarrow \bm{P}(\bm{I} - \bm{QD}) &= \bm{I} - \bm{H}' \nonumber\\
\Rightarrow \bm{P} - \bm{PQD} &= \bm{I} - \bm{H}'
\label{eqn:A.0.13}
\end{align}

The RHS of (\ref{eqn:A.0.13}) has all $1'$s in it's diagonal. Therefore the diagonal elements of $\bm{P}$ and $\bm{PQ}$ i.e. $\bm{(P)}_{i,i}$ and $\bm{(PQ)}_{i,i}$  cannot be simultaneously 0.  

\textbf{\underline{Case 1:}} $\bm{(P)}_{i,i} \neq 1$ for some $i$. 

Without loss of generality let, 
$\bm{P} = \begin{pmatrix}\bm{P_{1}} & \bm{0}\\ \bm{0} & \bm{I}\end{pmatrix}$ where $\bm{P_{1}}$ is the matrix that has 0 in it's diagonal. Clearly, $\bm{P_{1}} = \begin{pmatrix}\bm{I} & \bm{0}\end{pmatrix} \bm{P} \begin{pmatrix}\bm{I} \\ \bm{0}\end{pmatrix} $. Now from (\ref{eqn:A.0.13}), 

\begin{align}
\begin{pmatrix}\bm{I} & \bm{0}\end{pmatrix} (\bm{P - PQD}) \begin{pmatrix}\bm{I} \\ \bm{0}\end{pmatrix} &= \begin{pmatrix}\bm{I} & \bm{0}\end{pmatrix} (\bm{I - H}') \begin{pmatrix}\bm{I} \\ \bm{0}\end{pmatrix}\nonumber\\
\Rightarrow \bm{P_{1}} - \begin{pmatrix}\bm{P_{1}} & \bm{0}\end{pmatrix} \bm{Q} \begin{pmatrix}\bm{D_{1}} \\ \bm{0}\end{pmatrix} &= \bm{I} - \begin{pmatrix}\bm{I} & \bm{0}\end{pmatrix} \bm{H}' \begin{pmatrix}\bm{I} \\ \bm{0}\end{pmatrix}\nonumber\\
\Rightarrow  \bm{P_{1}} - \bm{P_{1}Q_{11}D_{1}} &= \bm{I} - \bm{H}'_{\bm{11}}
\label{eqn:A.0.14}      
\end{align}

Notice that, the diagonals of RHS of (\ref{eqn:A.0.14}) are equal to 1, $\bm{Q_{11}}$ is not necessarily a permutation matrix but it has at most one 1 in every column and $\bm{P_{1}}$ is a permutation matrix. Following from the same argument as before, we can therefore say that the diagonals of $\bm{P_{1}}$ and $\bm{P_{1}Q_{11}}$ cannot be simultaneously 0. Now from our assumption since $\bm{(P_{1})}_{i,i} = 0, \forall ~i$ we have, $$\bm{(P_{1}Q_{11})}_{i,i} = 1, \forall ~i$$
Now since $\bm{P_{1}}$ is a permutation matrix and $\bm{Q_{11}}$ has at most one 1 in every column, we have $\bm{P_{1}Q_{11} = I}$ and $\bm{D_{1} = -I}$
Therefore, from above we obtain,
\begin{align}
  \bm{I + P_{1}} &= \bm{I} - \bm{H_{11}}'\nonumber\\
  \Rightarrow \bm{P_{1}} &= -\bm{H_{11}}'
  \label{eqn:A.0.15}
\end{align}

Let any eigenvalue of matrix $\bm{A}$ be donoted by $\lambda(\bm{A})$. Therefore from (\ref{eqn:A.0.15}), we can obtain, 
\begin{align*}
    \lambda(\bm{P_{1}}) &= \lambda(-\bm{H_{11}}')\\
    \Rightarrow \lambda(\bm{P_{1}}) &= -\lambda(\bm{H_{11}}'), (\because \text{$-\lambda$ is an eigenvalue for $\bm{H_{11}}$})\\
    \Rightarrow |\lambda(\bm{P_{1}})| &= |\lambda(\bm{H_{11}}')|, (\text{taking modulus on both sides})
\end{align*}

Now since $\bm{P_{1}}$ is a permutation matrix, all of it's eigenvalues lie on a unit circle, i.e. $|\lambda(\bm{P_{1}})| = 1$. But according to Assumption \ref{ass3} of the main manuscript, the moduli of the eigenvalues of $\bm{H}'$ and hence $\bm{H_{11}}'$ are less than 1 and none of the real eigenvalues are equal to 1. Therefore, we arrive at a contradiction.

\textbf{\underline{Case 2:}} $\bm{(P)}_{i,i} = 0 ~\forall i$

Therefore, $\bm{(PQ)}_{i,i} = 1~\forall i$ and $\bm{D} = -\bm{I}$. Therefore from (\ref{eqn:A.0.13}), we obtain, 
$$\bm{P} + \bm{I} = \bm{I} - \bm{H}'$$
Proceeding similarly from the case 1 argument, we arrrive at a contradiction. 
$$\therefore \bm{P} = \bm{I}$$

\end{proof}

\section{Posterior inference}
\label{posterior}
\subsection{Selecting the effective number of basis functions for the causal embedded space}
While it is possible to use a prior to learn the number of basis functions jointly with other parameters through reversible jump MCMC or to use shrinkage priors to adaptively truncate and eliminate redundant functions, these approaches can lead to significant computational burden and potential Markov chain mixing issues. Therefore, this article employs a simple heuristic approach, as described in \citealp{kowal2017bayesian, fBN}. First, the functional observations are imputed and arranged into a $(n \times p) \times d$ matrix, where $d = |\cup_{i,j} \mathcal{T}^{(i)}_{j} |$ represents the size of the union of the measurement grid over all realized random functions. Then, singular value decomposition is performed, and the minimum value of $K$ is selected such that its proportion of variance explained is at least 90\%. This value is fixed throughout MCMC. It should be noted that while $K$ remains fixed, the basis functions are adaptively inferred.

We have noted that the value of $K$ derived from the aforementioned heuristic method falls within a range of $\pm{2}$ in comparison to the value obtained by fixing a grid encompassing values $\{1,2,3,4,5,6,7\}$ for $K$ and subsequently selecting the $K$ associated with the lowest WAIC \citep{waic}. The graph recovery performance, as assessed by Matthew's correlation coefficient (MCC) using this method, closely aligns with that of the previous approach. Consequently, we adopted the aforementioned heuristic technique to determine the optimal number of basis functions that collectively span the causal embedded space.

\subsection{Posterior distributions}
While the closed form expression for the posterior distribution cannot be obtained, we resort to MCMC techniques for sampling. We use superscript $(\cdot)$ to denote observations throughout the text. Let $\bm{X}^{(1)}, \dots, \bm{X}^{(n)}$ be $n$ realizations of the multivariate random functions $\bm{X}$. For the mixture of Gaussian distribution we assume $M_{jk} = M$ for simplicity. In order to obtain updates for the parameters of the mixture distribution, we define a class variable $\bm{\xi}^{(i)} = (\bm{\xi}^{(i)\top}_{1}, \dots, \bm{\xi}^{(i)\top}_{p}, \bar{\bm{\xi}}^{(i)\top}_{1}, \dots, \bar{\bm{\xi}}^{(i)\top}_{p})^{\top}$ with $\bm{\xi}^{(i)}_{j} = (\xi^{(i)}_{j1}, \dots, \xi^{(i)}_{jK_{j}})^{\top}$ and $\bar{\bm{\xi}}^{(i)}_{j} = (\xi^{(i)}_{j, K_{j}+1}, \dots, \xi^{(i)}_{jS})^{\top}$ where $\xi^{(i)}_{jk} = m$ if $\tilde{\epsilon}^{(i)}_{jk}$ belongs to the mixture component $m$. Let $\bm{M}^{(i)} = (\bm{\mu}^{(i)\top}_{1}, \dots, \bm{\mu}^{(i)\top}_{p}, \bar{\bm{\mu}}^{(i)\top}_{1}, \dots, \bar{\bm{\mu}}^{(i)\top}_{p})^{\top}$ and $\bm{T}^{(i)} = \text{diag}(\bm{\tau}^{(i)}_{1}, \dots, \bm{\tau}^{(i)}_{p}, \bar{\bm{\tau}}^{(i)}_{1}, \dots, \bar{\bm{\tau}}^{(i)}_{p})$ be the mean and covariance matrix of $\bm{\epsilon}^{(i)}$ where $\bm{\mu}^{(i)}_{j} = (\mu^{(i)}_{j1}, \dots, \mu^{(i)}_{jK_{j}})^{\top}$, $\bar{\bm{\mu}}^{(i)}_{j} = (\mu^{(i)}_{j,K_{j}+1}, \dots, \mu^{(i)}_{jS})^{\top}$, $\bm{\tau}^{(i)}_{j} = (\tau^{(i)}_{j1}, \dots, \tau^{(i)}_{jS})^{\top}$ and $\bar{\bm{\tau}}^{(i)}_{j} = (\tau^{(i)}_{j,K_{j}+1}, \dots, \tau^{(i)}_{jS})^{\top}$ with $\mu^{(i)}_{jk} = \sum_{m = 1}^{M}\mu_{jkm}\mathbf{1}(\xi^{(i)}_{jk} = m)$ and $\tau^{(i)}_{jk} = \sum_{m = 1}^{M}\tau_{jkm}\mathbf{1}(\xi^{(i)}_{jk} = m)$. Define $\bm{\pi}_{jk} = (\pi_{jk1}, \dots, \pi_{jkm})^{\top}, \forall j\in[p], k\in[S]$. Let $\tilde{\bm{\epsilon}}^{(i)} = \tilde{\bm{\alpha}}^{(i)} - \tilde{\bm{B}}\tilde{\bm{\alpha}}^{(i)}$ be the vector of exogenous variables for the $i^{\text{th}}$ observation.

\paragraph{Posterior distribution of the parameters of the mixture distribution.} For each $j \in [p], k \in [S]$, update the mixture weights $\bm{\pi}_{jk}$ by drawing from a Dirichlet distribution with concentration parameters $\{\beta_{m}\}_{m\in [M]}$ where, 

\begin{align}
    \beta_{m} &= \alpha + \sum_{i=1}^{n}\mathbf{1}(\xi^{(i)}_{jk} = m) \label{upd_mixwts}
\end{align}

Now, given the $\bm{\pi}_{jk}$'s, for each $i \in [n], j\in[p], k \in [S]$, update the class variables $\xi^{(i)}_{jk}$ from a categorical distribution with class probability $\{\pi^{(i)}_{m}\}_{m\in M}$ where, $\pi^{(i)}_{m} \propto \pi_{jkm}\text{N}(\tilde{\epsilon}^{(i)}_{jk}; \mu_{jkm}, \tau_{jkm})$ with $\sum_{m = 1}^{M}\pi^{(i)}_{m} = 1$. 

\begin{align}
    \pi^{(i)}_{m} \propto \pi_{jkm}\text{N}(\tilde{\epsilon}^{(i)}_{jk}; \mu_{jkm}, \tau_{jkm}),~~\sum_{m = 1}^{M}\pi^{(i)}_{m} = 1
    \label{upd_classlabels}
\end{align}

Next, for each $j\in[p], k \in [S], m\in [M]$ we update the mean parameter $\mu_{jkm}$ by sampling from a $\text{N}(p_{jkm}, q^{-1}_{jkm})$ distribution with, 

\begin{align}
    \begin{split}
        q_{jkm} &= \left(1/b_{\mu} + \sum_{i=1}^{n}\mathbf{1}(\xi^{(i)}_{jk} = m)\right)\\
        p_{jkm} &= q^{-1}_{jkm}\left(a_{\mu} + \sum_{i=1}^{n}\mathbf{1}(\xi^{(i)}_{jk} = m)\tilde{\epsilon}^{(i)}_{jk}\right)
    \end{split}
    \label{upd_meanpara}
\end{align}

The variance parameter $\tau_{jkm}$ by sampling from a $\text{IG}(p'_{jkm}, q'_{jkm})$ where, 

\begin{align}
    \begin{split}
        p'_{jkm} &= a_{\tau} + 1/2 \sum_{i=1}^{n} \mathbf{1}(\xi^{(i)}_{jk} = m)\\
        q'_{jkm} &= b_{\tau} + 1/2 \sum_{i=1}^{n} \mathbf{1}(\xi^{(i)}_{jk} = m)(\tilde{\epsilon}^{(i)}_{jk} - \mu_{jkm})^{2}
    \end{split}
    \label{upd_varpara}
\end{align}
 
\paragraph{Posterior distribution of the orthonormal basis coefficients:} For each $i\in[n]$, define $\bm{L}^{(i)} = (\bm{I} - \tilde{\bm{B}})^{\top}\bm{T}^{(i)-1}(\bm{I} - \tilde{\bm{B}})$, $\bm{D}^{(i)}_{1} = \text{diag}(\bm{D}^{(i)}_{11}, \dots, \bm{D}^{(i)}_{1p})$ with $\bm{D}^{(i)}_{1j} = \left(\sum_{t \in \mathcal{T}^{(i)}_{j}} \bm{\phi}(t)\bm{\phi}(t)^{\top} \middle/ \sigma_{j}\right)$ and $\bm{D}^{(i)}_{2} = (\bm{d}^{(i)\top}_{21}, \dots, \bm{d}^{(i)\top}_{2p})^{\top}$ with $\bm{d}^{(i)}_{2j} = \left(\sum_{t\in \mathcal{T}^{(i)}_{j}} X^{(i)}_{jt}\bm{\phi}(t) \middle/ \sigma_{j}\right)$. Now we sample $\tilde{\bm{\alpha}}^{(i)}$ from $\text{N}_{pS}(\bm{p}^{(i)}_{\alpha}, \bm{Q}^{(i)-1}_{\alpha})$ where,

\begin{align}
    \begin{split}
        \bm{Q}^{(i)}_{\alpha} &= 
(\bm{D}^{(i)}_{1} + \bm{L}^{(i)}) \\
     \bm{p}^{(i)}_{\alpha} &= \bm{Q}^{(i)-1}_{\alpha}\left(\bm{D}^{(i)}_{2} + (\bm{I} - \tilde{\bm{B}})'\bm{T}^{(i)-1}\bm{M}^{(i)}\right)
    \end{split} \label{upd_orthobasis}
\end{align}

\paragraph{Posterior distribution of the noise variances:} For each $j \in [p]$, update $\sigma_{j}$ by sampling from $\text{IG}\left(p_{\sigma}, q_{\sigma}\right)$ where, 

\begin{align}
    \begin{split}
        p_{\sigma} &= a_{\sigma} + 1/2\sum_{i=1}^{n}T^{(i)}_{j}\\
        q_{\sigma} &= b_{\sigma} + 1/2 \sum_{i=1}^{n}\sum_{t \in \mathcal{T}^{(i)}_{j}}\left(X^{(i)}_{jt} - \tilde{\bm{\alpha}}^{(i)\top}_{j}\bm{\phi}(t)\right)^{2}
    \end{split}
    \label{upd_noisevar}
\end{align}

\paragraph{Posterior distribution of the edge formation probability:} Update the edge probability $r$ by drawing from a $\text{Beta}(p_{r}, q_{r})$ distribution where, 

\begin{align}
    \begin{split}
        p_{r} &= a_{r} + \sum_{j\neq \ell } E_{j\ell}\\
        q_{r} &= b_{r} + \sum_{j\neq \ell} (1 - E_{j\ell})
    \end{split}
    \label{upd_edgeprob}
\end{align}

\paragraph{Posterior distribution of the causal effect size:} Update $\gamma$ by drawing from a $\text{IG}(p_{\gamma}, q_{\gamma})$ where, 

\begin{align}
    \begin{split}
        p_{\gamma} &= a_{\gamma} + K^{2}/2\sum_{j\neq \ell} E_{j\ell}\\
        q_{\gamma} &= b_{\gamma} + 1/2\sum_{j\neq \ell}E_{j\ell}~\text{trace}(\bm{B}_{j\ell}^{\top} \bm{B}_{j\ell})
    \end{split}
    \label{upd_causeffect}
\end{align}

\paragraph{Posterior distribution of the coefficients of the bspline coefficients:} Define for each $i \in [n], j\in[p], k\in[S]$, $\tilde{X}^{(i)}_{jt, -k} = X^{(i)}_{jt} - \sum_{\substack{h=1\\h \neq k}}^{S}\tilde{\alpha}^{(i)}_{jh}\phi_{h}(t)$.
For each $k \in [S]$, we draw $\bm{\tilde{A}}^{U}_{k}$ from $\text{N}_{R}(\bm{p}_{k}, \bm{Q}_{k})$ where, 

\begin{align}
    \begin{split}
        \bm{Q}_{k} &= \left[\left\{\sum_{i=1}^{n}\sum_{j=1}^{p}\frac{(\tilde{\alpha}^{(i)}_{jk})^{2}}{\sigma_{j}}\sum_{t\in \mathcal{T}^{(i)}_{j}}\bm{b}(t)\bm{b}(t)^{\top}\right\} + \bm{S}^{-1}_{k}\right]^{-1}\\
        \bm{p}_{k} &= \bm{Q}_{k} \left[\sum_{i=1}^{n}\sum_{j=1}^{p}\sum_{t\in \mathcal{T}^{(i)}_{j}}\frac{\tilde{\alpha}^{(i)}_{jk}}{\sigma_{j}}\tilde{X}^{(i)}_{jt, -k}\bm{b}(t)\right]
    \end{split}
    \label{upd_bsplinecoef}
\end{align}

Now, denote $\bm{P}_{k} = \bm{J}\bm{\tilde{A}}_{-k}$, where $\bm{J} = \int \bm{\tilde{b}}(\omega)\bm{\tilde{b}}^{\top}(\omega)\, d\omega$. Finally, transform and normalize the unconstrained sample to $\bm{\tilde{A}}_{k}^{N} = \bm{\tilde{A}}^{U}_{k} - \bm{Q}_{k}\bm{P}_{k}(\bm{P}^{\top}_{k}\bm{Q}_{k}\bm{P}_{k})^{-1}\bm{P}_{k}\bm{\tilde{A}}^{U}_{k}$ and $\bm{\tilde{A}}_{k} = \bm{\tilde{A}}^{N}_{k} \times ([\bm{\tilde{A}}^{N}_{k}]^{\top}\bm{J}\bm{\tilde{A}}^{N}_{k})^{-1/2}$.

\paragraph{Posterior distribution of the regularization parameter.} Independently for each $k \in [S]$, conditional on all other parameters, denote $q_{k} = 1/2 \sum_{r=3}^{R}\tilde{A}^{2}_{kr}$ and $p = R/2$. We then draw each $\lambda_{k}$ from a $\text{Gamma}(p, q_{k})$ distribution truncated at $(L_{k}, U_{k})$.

\paragraph{Posterior distribution of the adjacency and causal effect matrices.} Recursively for each $E_{j\ell}, j,\ell\in[p]$, we perform a birth/death move such that $\bm{E}' = \bm{E}$ except $E'_{j\ell} = 1 - E_{j\ell}$. The joint posterior of $(\bm{B}_{j\ell}, E_{j\ell})$ does not have closed form expression and therefore we perform a Metropolis Hastings (MH) step for joint acceptance or rejection of $(\bm{B}_{j\ell}, E_{j\ell})$. First we draw a $\bm{B}'_{j\ell}$ from a proposal distribution $N(\bm{B}_{j\ell}, z \bm{I}_{K}, \bm{I}_{K})$. We check whether $\bm{B}' (= \bm{B}$ except for $j, \ell$ block entry)  satisfies the eigenvalue condition given in Assumption \ref{ass2} of the main manuscript. If yes then we proceed to the next step and if not, we draw another $\bm{B}'_{j\ell}$ from the proposal ditribution. Here $z$ is a tuning parameter for the MH step. Next we calculate the acceptance ratio($\alpha$) $= \alpha_{N} - \alpha_{D}$ where,
\begin{multline}
    \alpha_{N} = E'_{j\ell}\log\left(r MVN(\bm{B}'_{j\ell}; \bm{B}_{j\ell}, \gamma \bm{I}_{K}, \bm{I}_{K})\right) + (1 - E'_{j\ell})\log\left((1-r)MVN(\bm{B}'_{j\ell}; \bm{B}_{j\ell}, s\gamma \bm{I}_{K}, \bm{I}_{K})\right) + \\  \sum_{i=1}^{n} \log\left(N(\tilde{\bm{\alpha}}^{(i)}; (\bm{I} - \tilde{\bm{B}}')^{-1}\bm{M}^{(i)}, (\bm{I} - \tilde{\bm{B}}')^{\top}\bm{T}^{(i)-1}(\bm{I} - \tilde{\bm{B}}'))\right) \label{alpha_num}
\end{multline}
\begin{multline}
    \alpha_{D} = E_{j\ell}\log\left(r MVN(\bm{B}_{j\ell}; \bm{B}'_{j\ell}, \gamma \bm{I}_{K}, \bm{I}_{K})\right) + (1 - E_{j\ell})\log\left((1-r)MVN(\bm{B}_{j\ell}; \bm{B}'_{j\ell}, s\gamma \bm{I}_{K}, \bm{I}_{K})\right) + \\  \sum_{i=1}^{n} \log\left(N(\tilde{\bm{\alpha}}^{(i)}; (\bm{I} - \tilde{\bm{B}})^{-1}\bm{M}^{(i)}, (\bm{I} - \tilde{\bm{B}})^{\top}\bm{T}^{(i)-1}(\bm{I} - \tilde{\bm{B}}))\right)
    \label{alpha_den}
\end{multline}
Then we accept or reject the proposed $(\bm{B}'_{j\ell}, E'_{j\ell})$ based on whether the value of a uniform random variable is less than or greater than $\min\{1, \alpha\}$. The value of $z$ is tuned to achieve an acceptance rate between $20\% ~\text{to}~ 40\%$.

\subsection{Markov Chain Monte Carlo algorithm} 
In this section we delineate the steps of Markov Chain Monte Carlo algorithm for drawing samples from the posterior distributions.

\begin{algorithm}
\caption{MCMC algorithm to obtain posterior samples} \label{mcmcalgo}
\begin{algorithmic}[1]
    \For{$b \gets 1$ to $B$}
       \For{$i \gets 1$ to $n$}
       \State Draw $\tilde{\bm{\alpha}}^{(i), [b]} \sim \text{N}_{pS}(\bm{p}^{(i)}_{\alpha}, (\bm{Q}^{(i)}_{\alpha})^{-1})$; \Comment{Update the basis coefficients by (\ref{upd_orthobasis})}
      \EndFor
    \EndFor
    
    \For{$j \gets 1$ to $p$}
         \State Draw $\sigma_{j}^{[b]}$ from $\text{IG}\left(p_{\sigma}, q_{\sigma}\right)$; \Comment{Update the noise variances by (\ref{upd_noisevar})}
    \EndFor
      
    \State Draw $r^{[b]}$ from $\text{Beta}(p_{r}, q_{r})$; \Comment{Update the edge formation probability by (\ref{upd_edgeprob})}

    \State Draw $\gamma^{[b]}$ from $\text{IG}(p_{\gamma}, q_{\gamma})$; \Comment{Update the causal effect size by (\ref{upd_causeffect})}

    \For{$k \gets 1$ to $S$}
        \State Draw $\bm{\tilde{A}}^{U}_{k} \sim \text{N}_{R}(\bm{p}_{k}, \bm{Q}_{k})$; \Comment{Update the un-normalized bspline coefficients by (\ref{upd_bsplinecoef})}
        \State Calculate $\bm{P}_{k} = \bm{J}\bm{\tilde{A}}_{-k}$;
        \State Calculate $\bm{\tilde{A}}_{k}^{N} = \bm{\tilde{A}}^{U}_{k} - \bm{Q}_{k}\bm{P}_{k}(\bm{P}^{\top}_{k}\bm{Q}_{k}\bm{P}_{k})^{-1}\bm{P}_{k}\bm{\tilde{A}}^{U}_{k}$;
        \State Normalize $\bm{\tilde{A}}_{k}^{[b]} = \bm{\tilde{A}}^{N}_{k} \times ([\bm{\tilde{A}}^{N}_{k}]^{\top}\bm{J}\bm{\tilde{A}}^{N}_{k})^{-1/2}$;

    \EndFor

    \For{$k \gets 1$ to $S$}
    \State Draw $\lambda_{k}^{[b]} \sim \text{Gamma}\left(\frac{R}{2},  \frac{\sum_{r=3}^{R}\left(\tilde{A}^{[b]}_{kr}\right)^{2}}{2}\right)$; \Comment{Update the regularization parameter}
    \EndFor

    \algstore{myalg}
    
    \end{algorithmic}
\end{algorithm}

\begin{algorithm}                     
\begin{algorithmic} [1]                   
\algrestore{myalg}
\For{$j \gets 1$ to $p$}
       \For{$k \gets 1$ to $K$}
         \State Draw $\bm{\pi}^{[b]}_{jk} \sim \text{Dir}(\beta_{1}, \dots, \beta_{M})$ \Comment{Update the mixing weights by (\ref{upd_mixwts})}
         \For{$i \gets 1$ to $n$}
            \State Draw $\bm{\xi}^{(i),[b]}_{jk} \sim \text{Cat}(\{\pi^{(i),[b]}_{m}\}_{m\in [M]})$; \Comment{Update the class labels by (\ref{upd_classlabels})}
         \EndFor
       \EndFor
    \EndFor

   \For{$j \gets 1$ to $p$}
       \For{$k \gets 1$ to $K$}
         \For{$m \gets 1$ to $M$}
            \State Draw $\mu^{[b]}_{jkm} \sim \text{N}(p_{jkm}, q^{-1}_{jkm})$; \Comment{Update the mean parameter by (\ref{upd_meanpara})}
            \State Draw $\tau^{[b]}_{jkm} \sim \text{IG}(p'_{jkm}, q'_{jkm})$;\Comment{Update the variance parameter by (\ref{upd_varpara})}
         \EndFor
       \EndFor
    \EndFor

    \For{$j \gets 1$ to $p$}
       \For{$\ell \gets 1$ to $p$}
        \State Update $(E_{j\ell}, \bm{B}_{j\ell})$ by MH step using (\ref{alpha_den}) and  (\ref{alpha_num})
       \EndFor
    \EndFor

\end{algorithmic}
\end{algorithm}

\section{Some additional simulations}
\label{additionalsimulations}
\subsection{Misspecification analysis of the proposed model} \label{misspecification}

\subsubsection{With general exogenous variable distributions}
In this section we consider simulating the exogenous variable from distributions other than that of laplace distribution and compare the performance of our algorithm. In particular, following \citealp{Shimizu2011DirectLiNGAMAD}, we generate the exogenous variable $\epsilon_{jk}$ from (1) Student t distribution with 1 degrees of freedom, (2) Uniform (3) Exponential, (4)  Mixture of two double exponentials, (5) Symmetric mixture of four Gaussians, and (6) Non symmetric mixture of two Gaussians. Across all exogenous variable distributions, Table \ref{tab:example2} shows that the proposed FENCE model had the best performance.

\begin{table}[htb]
\caption{Table showing comparison of several methods for different distributions of exogenous variables $\epsilon_{jk}$ under 50 replicates}
  \label{tab:example2}
  \resizebox{\columnwidth}{!}{\begin{tabular}{c|*3c|*3c|*3c|*3c|*3c}
    \toprule
    \multirow{2}{*}{Distributions} & \multicolumn{3}{c|}{FENCE}  & \multicolumn{3}{c|}{fLiNG} & \multicolumn{3}{c|}{fPCA-LINGAM} & \multicolumn{3}{c|}{fPCA-PC} & \multicolumn{3}{c}{fPCA-CCD}\\  \cmidrule(l){2-4} \cmidrule(l){5-7} \cmidrule(l){8-10} \cmidrule(l){11-13} \cmidrule(l){14-16}
     & TPR & FDR & MCC & TPR & FDR & MCC & TPR & FDR & MCC & TPR & FDR & MCC & TPR & FDR & MCC \\ \hline
    (1) & 0.81(0.04) & 0.24(0.07) & 0.76(0.05) & 0.71(0.09) & 0.69(0.05) & 0.36(0.04) & 0.85(0.02) & 0.84(0.04) & 0.28(0.08) & 0.81(0.05) & 0.71(0.06) & 0.26(0.07) & 0.91(0.02) & 0.62(0.04) & 0.38(0.02)\\
    (2) & 0.75(0.04) & 0.21(0.03) & 0.86(0.04) & 0.73(0.04) & 0.68(0.04) & 0.33(0.07) & 0.82(0.06) & 0.76(0.04) & 0.26(0.02) & 0.83(0.06) & 0.67(0.04) & 0.30(0.05) & 0.87(0.02) & 0.69(0.05) & 0.35(0.04) \\ 
    (3) & 0.77(0.04) & 0.23(0.05) & 0.83(0.04) & 0.74(0.04) & 0.63(0.03) & 0.32(0.04) &  0.86(0.04) & 0.81(0.03) & 0.24(0.03) & 0.81(0.03) & 0.76(0.03) & 0.31(0.03) & 0.89(0.02) & 0.73(0.05) & 0.41(0.03) \\ 
    (4) & 0.88(0.07) & 0.14(0.06) & 0.89(0.05) & 0.67(0.07) & 0.75(0.06) & 0.29(0.05) & 0.81(0.02) & 0.79(0.05) & 0.22(0.09) & 0.82(0.08) & 0.75(0.04) & 0.27(0.05) & 0.83(0.03) & 0.58(0.03) & 0.43(0.03)  \\
    (5) & 0.81(0.07) & 0.21(0.06) & 0.87(0.05) & 0.69(0.06) & 0.71(0.05) & 0.25(0.04) & 0.84(0.03) & 0.76(0.02) & 0.25(0.06) & 0.80(0.07) & 0.73(0.05) & 0.29(0.03) & 0.86(0.04) & 0.67(0.05) & 0.36(0.02) \\
    (6) & 0.79(0.06) & 0.24(0.05) & 0.81(0.04) & 0.70(0.04) & 0.71(0.06) & 0.28(0.05)  & 0.82
    (0.04)& 0.73(0.05) & 0.31(0.03) & 0.83(0.07) & 0.71(0.05) & 0.25(0.03) & 0.78(0.02) & 0.68(0.04) & 0.39(0.05) \\ 
    
    \bottomrule
  \end{tabular}}
\end{table}

\subsubsection{With functions observed on unevenly spaced grids}
In this experiment, we generated simulated data with $(n, p)$ values of either $(500, 20)$, $(500, 50)$, $(800, 20),$ or $(800, 50)$. Unlike the method used in Section \ref{simulation_study} of the main manuscript, we initially selected 250 points at random from the uniform distribution between $0$ and $1$ and defined this set as $D$. For each realization $i$ of function $j$, we randomly selected a subset $D^{(i)}_j$ of size $m^{(i)}_{j} = 20$ from $D$ to measure the function. We generated the causal graph, direct causal effect matrix, orthonormal basis functions, basis coefficient sequences, and observations in the same way as Section \ref{simulation_study} of the main manuscript. We conducted this scenario $50$ times and compared the results with those from fLiNG, fPCA-LiNGAM, fPCA-PC and fPCA-CCD. The results presented in Table \ref{tab:example3} demonstrate that FENCE is effective and superior to these other methods in learning directed cyclic graphs for general multivariate functional data.

\begin{table}[htb]
\caption{Comaprison of various methods under unevenly specified grids under 50 replicates}
  \label{tab:example3}
  \resizebox{\columnwidth}{!}{\begin{tabular}{cc|*3c|*3c|*3c|*3c|*3c}
    \toprule
    \multirow{2}{*}{n} & \multirow{2}{*}{p} & \multicolumn{3}{c|}{FENCE}  & \multicolumn{3}{c|}{fLiNG} & \multicolumn{3}{c|}{fPCA-LINGAM} & \multicolumn{3}{c|}{fPCA-PC} & \multicolumn{3}{c}{fPCA-CCD}\\  \cmidrule(l){3-5} \cmidrule(l){6-8} \cmidrule(l){9-11} \cmidrule(l){12-14} \cmidrule(l){15-17}
     &  & TPR & FDR & MCC & TPR & FDR & MCC & TPR & FDR & MCC & TPR & FDR & MCC & TPR & FDR & MCC \\ \hline
    500 & 20 & 0.86(0.03) & 0.19(0.05) & 0.89(0.05) & 0.46(0.09) & 0.73(0.05) & 0.32(0.03) & 0.25(0.02) & 0.82(0.04) & 0.17(0.04) & 0.21(0.04) & 0.83(0.04) & 0.19(0.07) & 0.56(0.02) & 0.62(0.05) & 0.32(0.06)\\
    500 & 50 & 0.79(0.04) & 0.24(0.06) & 0.84(0.04) & 0.37(0.04) & 0.79(0.06) & 0.29(0.05) & 0.23(0.04) & 0.87(0.03) & 0.15(0.02) & 0.17(0.06) & 0.85(0.04) & 0.17(0.05) & 0.51(0.05) & 0.67(0.03) & 0.27(0.01) \\ \midrule
    800 & 20 & 0.91(0.04) & 0.16(0.03) & 0.91(0.04) & 0.61(0.07) & 0.79(0.06) & 0.41(0.04) & 0.33(0.03) & 0.79(0.05) & 0.25(0.02) & 0.35(0.03) & 0.81(0.02) & 0.31(0.03) & 0.78(0.02) & 0.56(0.01) &  0.39(0.03)\\ 
    800 & 50 & 0.88(0.07) & 0.20(0.04) & 0.88(0.05) & 0.55(0.03) & 0.81(0.02) & 0.38(0.05) & 0.27(0.02) & 0.86(0.05) & 0.22(0.09) & 0.31(0.06) & 0.82(0.04) & 0.29(0.05) & 0.73(0.03) & 0.64(0.04) & 0.36(0.05) \\
   \bottomrule
  \end{tabular}}
\end{table}

\subsubsection{When the true graph is acyclic}
In this section we compared our method with the fLiNG method under the assumption that the true graph is acyclic. The entire simulation setting remains same as that of described in Section \ref{simulation_study} of the main manuscript except that the true graph was generated under acyclicity constraint. It is observed from Table \ref{tab:example4}  that under this assumption, fLiNG has superior performance against the proposed FENCE model. 

\begin{table}[htb]
\caption{Comparison of two methods when the true graph is acyclic under 50 replicates}
  \label{tab:example4}
  \resizebox{\columnwidth}{!}{\begin{tabular}{ccc|*3c|*3c}
    \toprule
    \multirow{2}{*}{n} & \multirow{2}{*}{p} & \multirow{2}{*}{d} & \multicolumn{3}{c|}{FENCE}  & \multicolumn{3}{c}{fLiNG} \\  \cmidrule(l){4-6} \cmidrule(l){7-9} 
     &  &  & TPR & FDR & MCC & TPR & FDR & MCC  \\ \hline
    150 & 30 & 125 & 0.81(0.03) & 0.29(0.05) & 0.85(0.05) & 0.83(0.05) & 0.21(0.05) & 0.91(0.04)\\
    150 & 60 & 125 & 0.79(0.04) & 0.32(0.06) & 0.81(0.02) & 0.82(0.03) & 0.24(0.06) & 0.87(0.05)\\ 
    150 & 30 & 250 & 0.67(0.05) & 0.34(0.06) & 0.79(0.04) & 0.81(0.04) & 0.23(0.06) & 0.82(0.05)\\ 
    150 & 60 & 250 & 0.64(0.04) & 0.36(0.03) & 0.74(0.04) & 0.78(0.05) & 0.26(0.06) & 0.79(0.05)\\ \midrule
    300 & 30 & 125 & 0.85(0.03) & 0.23(0.05) & 0.87(0.04) & 0.79(0.05) & 0.19(0.06) & 0.93(0.04)\\ 
    300 & 60 & 125 & 0.81(0.04) & 0.26(0.05) & 0.81(0.08) & 0.85(0.02) & 0.21(0.05) & 0.87(0.04)\\
    300 & 30 & 250 & 0.77(0.02) & 0.31(0.06) & 0.81(0.05) & 0.78(0.03) & 0.23(0.06) & 0.85(0.05)\\
    300 & 60 & 250 & 0.75(0.03) & 0.35(0.04) & 0.79(0.05) & 0.82(0.03) & 0.24(0.05) & 0.80(0.05)\\
   \bottomrule
  \end{tabular}}
\end{table}

\subsubsection{When the true structural equation model is non-linear}
In this section, we have outlined the misspecification analysis for our model by generating data corresponding to a non-linear structural equation model (SEM). We have considered a scenario with number of samples$(n) = 100$, number of nodes$(p) = 6$  and evenly spaced time grid$(d)$ over $(0, 1)$ of size $d = 100$. The summary measures corresponding to $10$ replicates are given in Table \ref{tab:example5} below. The poor performance is clearly expected because our modeling assumptions involve linear SEM.

\begin{table}[htb]
\caption{Performance of FENCE when the true SEM is non-linear}
\label{tab:example5}
\centering
    \begin{tabular}{c|c|c}
    \toprule
    \multicolumn{3}{c}{FENCE} \\\hline
     TPR & FDR & MCC \\ \hline
     0.27(0.08) & 0.71(0.07) & 0.34(0.07)\\
     \bottomrule
    \end{tabular}
\end{table}

\subsection{Sensitivity analysis}
In this section, we outline how sensitive the performance of our model is against different choices of hyperparameters. The hyperparameters for our model are $(a_{\gamma}, b_{\gamma}), (a_{\tau}, b_{\tau}), (a_{\sigma}, b_{\sigma}),$ $ s, R, S, M$ and $\beta$. The data were generated the same way as in Section \ref{simulation_study} of the main manuscript with $(n, p, d) = (150, 20, 125)$. From Table \ref{tab:example6} we can conclude that the performance of our model is quite robust under different choice of hyperparameters. 

\begin{table}[htb]
\caption{Sensitivity analysis for different choices of hyperparameters. The metrics reported are based on 50 repetitions are reported; standard deviations are given within the
parentheses.}
    \label{tab:example6}
    \centering
    \resizebox{\columnwidth}{!}{\begin{tabular}{|c|c|c|c|c|c|c|c|c|c|}
    \toprule
       Hyperparameters & $(a_{\tau}, b_{\tau}) = (0.1, 0.1)$ & $(a_{\sigma}, b_{\sigma}) = (0.1, 0.1)$ & $(a_{\gamma}, b_{\gamma}) = (0.1, 0.1)$ & $s = 0.01$ & $R = 30$ & $S = 15$ & $M = 15$ & $\beta = 0.1$\\ \hline
        TPR & 0.79(0.02) & 0.80(0.02) & 0.78(0.03) & 0.75(0.03) & 0.79(0.02) & 0.80(0.01) & 0.81(0.02) & 0.82(0.02) \\
        FDR & 0.16(0.03) & 0.18(0.03) & 0.18(0.05) & 0.20(0.04) & 0.23(0.02) & 0.19(0.03) & 0.15(0.03) & 0.21(0.02) \\
        MCC & 0.76(0.04) & 0.81(0.04) & 0.83(0.04) & 0.82(0.03) & 0.81(0.01) & 0.84(0.04) & 0.83(0.01) & 0.84(0.03)\\
         \midrule
        Hyperparameters & $(a_{\tau}, b_{\tau}) = (0.1, 1)$ & $(a_{\sigma}, b_{\sigma}) = (0.01, 0.01)$ & $(a_{\gamma}, b_{\gamma}) = (0.1, 1)$ & $s = 0.03$ & $R = 20$ & $S = 10$ & $M = 20$ & $\beta = 2$\\
       \hline
        TPR & 0.80(0.02) & 0.79(0.03) & 0.76(0.02) & 0.75(0.03) & 0.78(0.03) & 0.79(0.02) & 0.82(0.04) & 0.81(0.03)\\
        FDR & 0.17(0.03) & 0.18(0.04) & 0.15(0.03) & 0.19(0.04) & 0.22(0.03) & 0.21(0.03) & 0.15(0.03) & 0.23(0.03) \\
        MCC & 0.77(0.02) & 0.80(0.02) & 0.82(0.02) & 0.82(0.03) & 0.80(0.02) & 0.83(0.03) & 0.83(0.03) & 0.85(0.05)\\
        \midrule
        Hyperparameters & $(a_{\tau}, b_{\tau}) = (5, 1)$ & $(a_{\sigma}, b_{\sigma}) = (0.1, 1)$ & $(a_{\gamma}, b_{\gamma}) = (5, 1)$ & $s = 0.05$ & $R = 25$ & $S = 20$ & $M = 30$ & $\beta = 5$\\
       \hline
        TPR & 0.76(0.04) & 0.82(0.05) & 0.79(0.03) & 0.76(0.04) & 0.78(0.03) & 0.80(0.03) & 0.81(0.02) & 0.79(0.04)\\
        FDR & 0.17(0.05) & 0.19(0.04) & 0.19(0.02) & 0.20(0.03) & 0.23(0.03) & 0.22(0.04) & 0.16(0.03) & 0.21(0.03)\\
        MCC & 0.78(0.02 )& 0.83(0.03) & 0.83(0.05) & 0.81(0.03) & 0.81(0.01) & 0.84(0.01) & 0.82(0.02) & 0.83(0.03)\\
        \bottomrule
    \end{tabular}}
\end{table}

\section{Comparison of various methods}\label{fulltable}
In this section, as discussed in Section \ref{simulation_study} of the main manuscript, we give the full summary of the simulation results related to the comparison of our method, FENCE, against fLiNG, fPCA-LiNGAM, fPCA-PC and fPCA-CCD in Table \ref{tab:example7}. Our conclusions remain the same. 

\begin{table}[htb]
\caption{Comparison of performance of various methods under 50 replicates. Since LiNGAM is not applicable to cases where $q > n$ with $q = Kp$ being the total number of extracted basis coefficients across all functions, the results from
those cases are not available and indicated by "-".}
\centering
  \label{tab:example7}
  \begin{adjustbox}{angle = 0}
  \resizebox{\columnwidth}{!}{\begin{tabular}{ccc|*3c|*3c|*3c|*3c|*3c}
    \toprule
    \multirow{2}{*}{n}& \multirow{2}{*}{p} & \multirow{2}{*}{d} & \multicolumn{3}{c|}{FENCE}  & \multicolumn{3}{c|}{fLiNG} & \multicolumn{3}{c|}{fPCA-LINGAM} & \multicolumn{3}{c|}{fPCA-PC} & \multicolumn{3}{c}{fPCA-CCD}\\  \cmidrule(l){4-6} \cmidrule(l){7-9} \cmidrule(l){10-12} \cmidrule(l){13-15} \cmidrule(l){16-18}
     &  &  & TPR & FDR & MCC & TPR & FDR & MCC & TPR & FDR & MCC & TPR & FDR & MCC & TPR & FDR & MCC \\ \hline
    75 & 20 & 125 & \textbf{0.85(0.09)} & \textbf{0.19(0.07)} & \textbf{0.88(0.05)} & 0.41(0.09) & 0.79(0.05) & 0.36(0.04) & 0.35(0.19) & 0.84(0.04) & 0.11(0.08) & 0.20(0.09) & 0.83(0.06) & 0.10(0.07) & 0.69(0.03) & 0.41(0.04) & 0.23(0.03)\\
    75 & 40 & 125 & \textbf{0.79(0.08)} & \textbf{0.23(0.06)} & \textbf{0.86(0.04)} & 0.37(0.08) & 0.82(0.06) & 0.33(0.05) & - & - & - & 0.11(0.06) & 0.91(0.04) & 0.05(0.05) & 0.73(0.02) & 0.47(0.04)& 0.21(0.05)\\ 
    75 & 60 & 125 & \textbf{0.75(0.07)} & \textbf{0.27(0.05)} & \textbf{0.83(0.04)} & 0.34(0.07) & 0.83(0.06) & 0.32(0.04) &  - & - & - & 0.11(0.03) & 0.91(0.03) & 0.06(0.03) & 0.68(0.03) & 0.61(0.05) & 0.19(0.03)\\ \midrule
    150 & 20 & 125 & \textbf{0.88(0.07)} & \textbf{0.14(0.06)} & \textbf{0.89(0.05)} & 0.45(0.07) & 0.75(0.06) & 0.39(0.05) & 0.28(0.22) & 0.86(0.05) & 0.08(0.09) & 0.31(0.08) & 0.75(0.04) & 0.12(0.05) & 0.71(0.03) & 0.42(0.03) & 0.25(0.04)\\
    150 & 40 & 125 & \textbf{0.81(0.07)} & \textbf{0.21(0.06)} & \textbf{0.87(0.05)} & 0.39(0.06) & 0.79(0.05) & 0.37(0.04) & 0.35(0.22) & 0.91(0.02) & 0.08(0.06) & 0.25(0.07) & 0.81(0.05) & 0.06(0.03) & 0.73(0.04) & 0.47(0.05) & 0.23(0.03)\\
    150 & 60 & 125 & \textbf{0.79(0.06)} & \textbf{0.24(0.05)} & \textbf{0.86(0.04)} & 0.36(0.04) & 0.80(0.06) & 0.36(0.05)  & - & - & - & 0.23(0.07) & 0.83(0.05) & 0.05(0.03) & 0.72(0.05) & 0.54(0.04) & 0.22(0.02)\\ \midrule
    300 & 20 & 125 & \textbf{0.91(0.03)} & \textbf{0.09(0.04)} & \textbf{0.90(0.04)} & 0.51(0.04) & 0.73(0.06) & 0.41(0.04) & 0.30(0.19) & 0.84(0.05) & 0.11(0.09) & 0.36(0.09) & 0.72(0.05) & 0.14(0.05) & 0.81(0.03) & 0.39(0.04) & 0.26(0.03)\\
    300 & 40 & 125 & \textbf{0.87(0.04)} & \textbf{0.15(0.05)} & \textbf{0.87(0.05)} & 0.47(0.05) & 0.75(0.06) & 0.38(0.05) & 0.27(0.20) & 0.91(0.02) & 0.08(0.06) & 0.29(0.06) & 0.76(0.06) & 0.07(0.03) & 0.77(0.03) & 0.45(0.02) & 0.24(0.03)\\
    300 & 60 & 125 & \textbf{0.85(0.05)} & \textbf{0.17(0.03)} & \textbf{0.86(0.03)} & 0.45(0.05) & 0.76(0.04) & 0.38(0.03) &  0.28(0.17) & 0.91(0.05) & 0.07(0.06) & 0.28(0.04) & 0.77(0.05) & 0.05(0.03) & 0.72(0.03) & 0.49(0.02) & 0.22(0.03)\\
    \midrule
    75 & 20 & 250 & \textbf{0.81(0.04)} & \textbf{0.23(0.02)} & \textbf{0.85(0.05)} & 0.39(0.07) & 0.80(0.05) & 0.39(0.04) & 0.32(0.14) & 0.82(0.03) & 0.09(0.04) & 0.19(0.07) & 0.81(0.04) & 0.13(0.07) & 0.67(0.03) & 0.46(0.03) & 0.22(0.04) \\
    75 & 40 & 250 & \textbf{0.73(0.04)} & \textbf{0.28(0.05)} & \textbf{0.82(0.04)} & 0.35(0.04) & 0.85(0.06) & 0.33(0.05) & - & - & - & 0.25(0.06) & 0.83(0.04) & 0.12(0.04) & 0.68(0.02) & 0.51(0.04) & 0.21(0.03)\\ 
    75 & 60 & 250 & \textbf{0.67(0.03)} & \textbf{0.34(0.05)} & \textbf{0.79(0.04)} & 0.34(0.04) & 0.85(0.03) & 0.31(0.04) &  - & - & - & 0.17(0.03) & 0.83(0.02) & 0.09(0.03) & 0.63(0.04) & 0.56(0.04) & 0.19(0.04)\\ \midrule
    150 & 20 & 250 & \textbf{0.83(0.06)} & \textbf{0.17(0.05)} & \textbf{0.86(0.05)} & 0.46(0.07) & 0.73(0.07) & 0.42(0.05) & 0.32(0.19) & 0.79(0.05) & 0.13(0.05) & 0.41(0.08) & 0.72(0.04) & 0.19(0.05) & 0.73(0.04) & 0.43(0.02) & 0.24(0.03)\\
    150 & 40 & 250 & \textbf{0.79(0.02)} & \textbf{0.26(0.06)} & \textbf{0.82(0.03)} & 0.41(0.05) & 0.71(0.05) & 0.40(0.03) & 0.31(0.14) & 0.81(0.02) & 0.13(0.06) & 0.46(0.07) & 0.73(0.05) & 0.15(0.02) & 0.71(0.03) & 0.47(0.04) & 0.23(0.03)\\
    150 & 60 & 250 & \textbf{0.69(0.05)} & \textbf{0.31(0.05)} & \textbf{0.79(0.04)} & 0.43(0.03) & 0.79(0.06) & 0.43(0.05)  & - & - & - & 0.41(0.03) & 0.75(0.05) & 0.14(0.03) & 0.69(0.02) & 0.52(0.03) & 0.21(0.02)\\ \midrule
    300 & 20 & 250 & \textbf{0.86(0.02)} & \textbf{0.16(0.04)} & \textbf{0.85(0.04)} & 0.68(0.02) & 0.77(0.07) & 0.47(0.04) & 0.45(0.13) & 0.86(0.05) & 0.17(0.09) & 0.42(0.09) & 0.86(0.03) & 0.13(0.05) & 0.78(0.02) & 0.44(0.06) & 0.27(0.03)\\
    300 & 40 & 250 & \textbf{0.79(0.08)} & \textbf{0.16(0.05)} & \textbf{0.84(0.06)} & 0.73(0.05) & 0.71(0.06) & 0.43(0.05) & 0.39(0.16) & 0.87(0.05) & 0.16(0.07) & 0.45(0.06) & 0.81(0.06) & 0.12(0.06) & 0.76(0.05) & 0.49(0.06) & 0.23(0.05)\\
    300 & 60 & 250 & 0.76(0.05) & \textbf{0.21(0.03)} & \textbf{0.80(0.03)} & \textbf{0.77(0.05)} & 0.74(0.03) & 0.42(0.03) &  0.28(0.17) & 0.90(0.04) & 0.13(0.04) & 0.43(0.04) & 0.79(0.07) & 0.12(0.04) & 0.72(0.06) & 0.53(0.03) & 0.22(0.04)\\
    \bottomrule
  \end{tabular}}
 \end{adjustbox}
\end{table}

\end{document}